\documentclass{amsart}

\makeatletter
\renewcommand{\email}[2][]{%
  \ifx\emails\@empty\relax\else{\g@addto@macro\emails{,\space}}\fi%
  \@ifnotempty{#1}{\g@addto@macro\emails{\textrm{(#1)}\space}}%
  \g@addto@macro\emails{#2}%
}
\makeatother

\usepackage{graphicx,epsfig,epstopdf,dsfont,color}\usepackage{soul,dcolumn,chngcntr,,tikz}
\usepackage{amsmath,amssymb,caption}
\usepackage{bm,gensymb,floatrow,subfig}
\usepackage{multirow}

\usepackage{graphicx,epsfig,epstopdf,dsfont,color}\usepackage{soul,dcolumn,chngcntr,,tikz}
\usepackage{caption}
\usepackage{bm,gensymb,floatrow,subfig}
\usepackage{multirow}

\usepackage{floatrow,subfig,amsmath,amssymb,dcolumn,bm,chngcntr,tikz}
\usepackage{amsaddr}
\usepackage{natbib}

\usepackage{upgreek}

\definecolor{darkblue}{RGB}{83,0,93}

\def\bes{\begin{eqnarray}}
\def\ees{\end{eqnarray}}
\def\beq{\begin{equation}}
\def\eeq{\end{equation}}
\def\ba#1\ea{\begin{align}#1\end{align}}
\def\bsa#1#2\esa{\begin{subequations}\label{#1}
\begin{align}#2\end{align} \end{subequations}}

\def\lb{\left[}
\def\rb{\right]}
\def\lcb{\left\{}
\def\rcb{\right\}}

\def\f{\frac}
\def\b{\mathbf}
\def\p{\partial}
\def\co{{\mathcal O}}

\def\u{\bm{u}}

\def\NA{\bm{\nabla}}
\def\DEL{\nabla^2}
\def\mt{\overline{\langle T \rangle}_x}
\def\mr{\overline{\langle \rho \rangle}_x}
\def\Nu{\text{Nu}}
\def\Ra{\text{Ra}}
\def\Pr{\text{Pr}}
\def\s{\text{S}}
\def\Z{Z_{\text{NB}}}

\numberwithin{equation}{section}
\usepackage[margin=1in]{geometry}

\begin{document}

\title{Dynamics of Mixed Convective--Stably-Stratified Fluids}

\author{L.-A. Couston$^1$, D. Lecoanet$^{2,3,1,4}$, B. Favier$^1$, M. Le Bars$^1$}
\address{$^1$ CNRS, Aix Marseille Univ, Centrale Marseille, IRPHE, Marseille, France \\ $^2$ Princeton Center for Theoretical Science, Princeton, NJ 08544, USA \\ $^3$ Department of Astrophysical Sciences, Princeton University, Princeton, NJ 08544, USA \\ $^4$ Kavli Institute for Theoretical Physics, University of California, Santa Barbara, CA 93106, USA}
\thanks{couston@irphe.univ-mrs.fr}

%\subjclass[2010]{Primary }


\date{}

\dedicatory{}

\begin{abstract}

We study the dynamical regimes of a density-stratified fluid confined between isothermal no-slip top and bottom boundaries (at temperatures $T_t$ and $T_b$) via direct numerical simulation.  The thermal expansion coefficient of the fluid is temperature dependent and chosen such that the fluid density is maximum at the inversion temperature $T_b>T_i>T_t$. Thus, the lower layer of the fluid is convectively unstable while the upper layer is stably stratified. We show that the characteristics of the convection change significantly depending on the degree of stratification of the stable layer. For strong stable stratification, the convection zone coincides with the fraction of the fluid that is convectively unstable (i.e. where $T>T_i$), and convective motions consist of rising and sinking plumes of large density anomaly, as is the case in canonical Rayleigh-B\' enard convection; internal gravity waves are generated by turbulent fluctuations in the convective layer and propagate in the upper layer. For weak stable stratification, we demonstrate that a large fraction of the stable fluid (i.e. with temperature $T<T_i$) is instead destabilized and entrained by buoyant plumes emitted from the bottom boundary. The convection thus mixes cold patches of low density-anomaly fluid with hot upward plumes, and the end result is that the $T_i$ isotherm sinks within the bottom boundary layer and that the convection is entrainment-dominated. We provide a phenomenological description of the transition between the regimes of plume-dominated and entrainment-dominated convection through analysis of the differences in the heat transfer mechanisms, kinetic energy density spectra, and probability density functions for different stratification strengths. Importantly, we find that the effect of the stable layer on the convection decreases only weakly with increasing stratification strength, meaning that the dynamics of the stable layer and convection should be studied self-consistently in a wide range of applications.

\end{abstract}

%\pacs{}

\maketitle

%%%%%%%%%%%%%%%%%%%%%%
\section{Introduction}
%%%%%%%%%%%%%%%%%%%%%%

Buoyancy-driven convection can result from a number of physical mechanisms leading to a change 
of the fluid density, including evaporation, heating, and sedimentation. Convection occurs naturally in the interior of stars and planets, as well as in oceans and atmospheres, but is also sometimes used to improve heat transfer properties in buildings and industrial processes. The transport of mass, momentum and heat are all important aspects of convective flows.

Because density variations are often small compared to the mean fluid density, buoyancy-driven convection studies typically neglect density effects other than in the buoyancy force, an approximation known as the Boussinesq assumption. The Oberbeck-Boussinesq assumption (OB), which additionally assumes constant physical properties (such as viscosity, conductivity, and thermal expansion), is also often considered and has resulted in much of our fundamental understanding of convective flows, spear-headed by the canonical problem of (thermally-driven) Rayleigh-B\' enard convection (RBC) \cite[][]{Ahlers2009}.

While the Boussinesq assumption is applicable in a number of physically-interesting fluid systems, some dynamics of realistic convective fluids are not attained under the Oberbeck-Boussinesq approximation. In thermal convection, for instance, symmetric top and bottom boundary conditions necessarily result in a top/down mid-plane symmetry of the flow under OB assumption. Helium convection experiments have demonstrated asymmetric fluid motions with respect to the mid-plane due to temperature-dependent viscosity \cite[][]{Urban2012}; an effect discarded within the OB framework. Non-Oberbeck-Boussinesq (NOB) effects are of current research interest because a number of fluids (e.g. water, helium, ethane) have temperature-dependent physical properties, in particular close to their critical point \cite[such as in glycerol, see e.g.][]{Horn2013}.

Of all NOB effects, those due to temperature-dependent expansion coefficients may be most striking \cite[in ethane, see e.g.][]{Ahlers2008}, and at the same time can be readily observed in nature. This is especially true because the expansion coefficient of water varies significantly close to its density maximum at temperature $T_0=4\degree$C, a temperature ubiquitous in oceans and lakes which can thus routinely experience NOB effects \cite[][]{Nycander2015,Vettoretti2016}. The density maximum for water results from the quadratic equation of state (EoS) $\rho\propto \rho_0[1-\beta(T-T_0)^2]$, which is a simplified yet realistic EoS for water at atmospheric pressure \cite[][]{Roquet2015}. Because of the quadratic temperature nonlinearity, the water with a positive vertical temperature gradient reverses from buoyancy decreasing with height for $T\leq T_0=4\degree C$ to buoyancy increasing with height for $T\geq T_0=4\degree C$. This results in a layer subject to convective instability beneath a stably stratified layer. The existence of an adjacent stable stratification can have a significant influence on the convective dynamic, most notably allowing for penetrative convection, as demonstrated theoretically via stability analysis decades ago \cite[e.g.][]{Veronis1963}. More recently, laboratory experiments \cite[][]{Lebars2015} and numerical simulations \cite[][]{Lecoanet2015} have improved our understanding of such penetrative convection following the seminal study by \cite{Townsend1964} and have shown a clear picture of plumes interacting with a stable layer and generating internal waves, an effect inaccessible within the OB framework.

NOB convection with a buoyancy reversal is not limited to water but also occurs in a number of alcohol-glycol--water mixtures used as analogues of gas dissolution in porous media \cite[][]{Backhaus2011,Hidalgo2012,Hewitt2013}, glass-forming liquids such as BeF$_2$ \cite[][]{Hemmati2001}, and in the troposphere where an inversion layer forms atop the cloud-top mixing layer as a result of evaporation \cite[][]{Mellado2010}. The heterogeneous convective/wave dynamic is also relevant to many astrophysical and geophysical settings: In stars such as our Sun, convection only occurs in the relatively cool outer shell, and couples with an underlying stable region where gravity waves propagate \cite[][]{Brummell2002,Rogers2005,Alvan2013,Pincon2016}; In the Earth, observations within the convective liquid outer core point to a possible stratified layer near both the core-mantle boundary and the inner-core boundary \cite[][]{Hirose2013,Buffett2014}. Buoyancy reversal in astrophysical and geophysical systems is due to changes from super- to sub-adiabatic mean temperature profiles, related to changes of physical properties with temperature and pressure.

In addition to the effects of the Rayleigh and Prandtl numbers on thermal convection, buoyancy reversal dynamics are strongly affected by an additional physical parameter, the relative strength ($\s$) of the stable stratification compared to the destabilizing buoyancy difference. A number of questions related to buoyancy reversal effects in thermal convection are related to the effect of varying $\s$. Such questions, addressed in this paper, include: for which $\s$ is it safe to assume that the convection is not affected by the dynamic of the stable layer? How is the generation of internal waves affected by $\s$? How does the height of the convection zone change with $\s$? Is the heat transfer sensitive to $\s$?

The idea of possibly decoupling the convection zone from the stable fluid layer is of particular interest because the convective and internal gravity wave dynamics can be on vastly different time and length scales \cite[][]{Lebars2015,Lecoanet2015}. In numerical simulations it would be advantageous to run a turbulent convection-only simulation, and then use the output flow statistics from this simulation as a source term for generating internal waves in a separate run. This idea was put forward theoretically by \cite{Goldreich1990}: assuming a Kolmogorov-type spectrum representative of Reynolds stresses due to sweeping plumes below the unstable/stable interface, they bypassed the simulation of the convection and were able to directly analyze in details the generated internal wavefield. Of course this approach requires a careful investigation of whether a Kolmogorov spectrum is appropriate and whether Reynolds stress is indeed the main generating mechanism.  Ongoing work and recent results on water convection seem to support that waves are mainly generated by Reynolds stresses, at least for some (high) values of $\s$ \cite[][]{Lecoanet2015}. Another simplified approach consists in deriving one-dimensional (along the vertical) models with turbulence diffusivity parameterization, as done in, for instance, \cite{Wunsch2003}.

Here we address via direct numerical simulations some of the questions pertaining to the effect of the stable stratification strength on buoyancy reversal convection. Because our goal is to obtain a clear picture of the various convective dynamics for a wide range of stratification parameters, we focus in this paper on two-dimensional simulations. Reducing the system to two dimensions allows us to run a large number of simulations for several diffusive times, which is necessary to reach a statistically steady state. We describe the model setup in \S\ref{sec2}, and in \S\ref{sectransients} we analyze the onset and transition to steady state. The qualitative features of simulations with high, moderate, and small stratification strengths are then described in \S\ref{sec3}. A detailed analysis is presented in \S\ref{sec4} and concluding remarks are offered in \S\ref{sec5}.

%%%%%%%%%%%%%%%%%%%%%%
\section{Problem formulation}\label{sec2}
%%%%%%%%%%%%%%%%%%%%%%

We consider the Navier-Stokes equations under Boussinesq approximation and in two dimensions. We take $x$ ($z$) to be the horizontal (vertical upward) direction, with $\hat{x}$ and $\hat{z}$ the unit vectors. The dimensionless equations are
\bsa{a3}\label{a31}
\p_t \u + \u\cdot\NA\u & = -\NA p + \Pr\DEL\u - \Pr \Ra \rho \hat{z} , \\ \label{a32}
\p_t T + \u\cdot\NA T & = \DEL T, \\ \label{a33}
\NA\cdot \u & = 0.
\esa
Length and time scales are non-dimensionalized using the full vertical extent of the domain $\ell$ and the thermal diffusion time $\ell^2/\kappa$, where $\kappa$ is the constant thermal diffusivity of the fluid. $T$ is the temperature, $\u$ the velocity vector, and $\rho$ is the density anomaly compared to the reference density $\rho_0$. We assume isothermal and no-slip boundary conditions on the top and bottom horizontal plates and periodicity of all fluid variables in the horizontal $x$ direction. The dimensionless parameter $\Pr=\nu/\kappa$ is the Prandtl number, with $\nu$ the constant fluid viscosity. $\Ra$ is the Rayleigh number, defined as  
\ba{}\label{a44}
\Ra=\f{\alpha_c^* g (T_b^*-T_i^*) \ell^3}{\nu\kappa},
\ea
where $g$ is the constant gravitational acceleration, $\alpha_c^*$ is the (dimensional) constant thermal expansion coefficient for temperatures $T^*>T_i^*$, and $T_b^*-T_i^*>0$ is the (dimensional) temperature scale obtained from the difference between the temperature on the bottom plate ($T_b^*$) and inversion temperature ($T_i^*$ defined below). Using $T_i^*$ as the absolute reference temperature then gives the dimensionless temperature $T$ in \eqref{a3} in terms of the dimensional one, $T^*$, as $T=(T^*-T_i^*)/(T_b^*-T_i^*)$, implying $T_b=1$ and $T_i=0$.

\begin{figure}
\centering
%\hspace{-0in}
\includegraphics[width=0.5\textwidth]{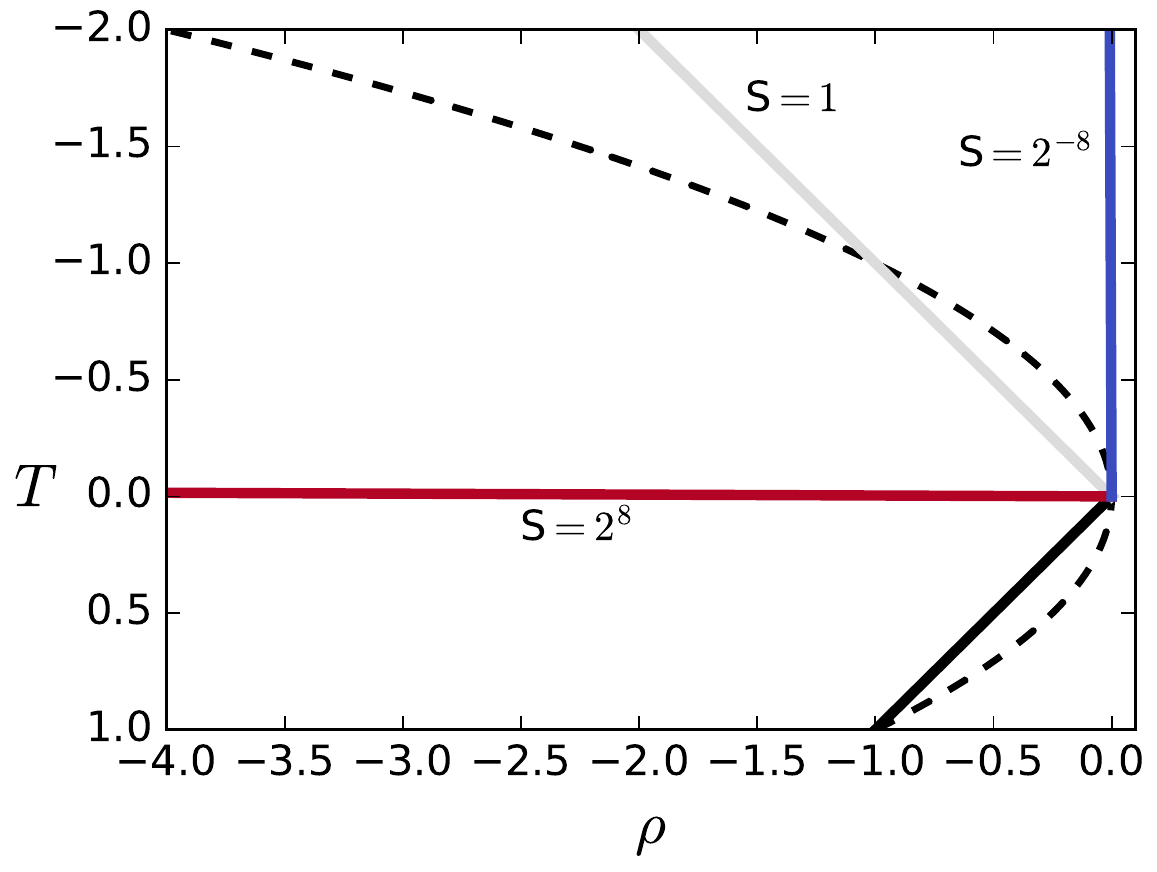}
%\vspace{-0.1in}
\caption{Equations of state for the density anomaly $\rho$ as a function of temperature $T$ ($T$ decreasing upward) for three different stiffness parameters $\s$ (solid lines). For comparison with water, the quadratic equation of state $\rho=-T^2$ is also shown (dashed line); note that the $\s$ parameter that best represents water depends on the temperature range considered.}\label{EoS}
\end{figure}

As in classical RBC, we consider an equation of state for the density anomaly of the form $\rho=-\alpha T$. However, we deviate from the OB paradigm as we allow the dimensionless thermal expansion coefficient $\alpha(T)$ to vary with temperature, as is the case for e.g. water. Specifically, as can be seen in figure \ref{EoS}, we consider 
\ba{}\label{aS}
\rho = -\alpha(T)T=\lcb \begin{array}{c}
-T,~T\geq 0, \\
\s T,~T< 0,
\end{array}\right.
\ea
such that $\alpha$ changes sign at $T=0$, and with $\s>0$ a free parameter of our model. The change of sign of $\alpha$ results in the density anomaly being non-monotonic and maximum at $T=0$, which we thus refer to as the inversion temperature (at the dimensional temperature $T_i^*$). The form of $\alpha$ is the simplest model that can reproduce a non-monotonic EoS similar to water (shown in figure \ref{EoS}), and that allows varying degrees of stratification of the top stable layer. While $\s\gg 1$ is obtained when considering a stronger stratification than is the case for water, i.e. a steeper effect of varying temperature on the density, $\s\ll 1$ implies a weaker stratification. It can be noted that a smoothed piecewise profile for $\alpha$ may have worked equally well as the discontinuous profile considered in equation \eqref{aS}, though it would have introduced additional parameters in the problem. Numerically, equation \eqref{aS} did not pose any problem, although a relatively high vertical resolution was necessary, because even though $\alpha$ is discontinuous, relevant terms appearing in the equations (i.e. $T$ and its first-order and second-order derivatives as well as the buoyancy $\alpha T$) are all continuous. The variations of $\rho$ with $T$ are shown in figure \ref{EoS} for $\s=1$, along with the stiffest $\s=2^{8}$ (almost horizontal line) and most flexible $\s=2^{-8}$ (almost vertical line) cases considered in this paper. For $\s=2^{8}$, note that the stratification in density is so strong in the stable layer that the density anomaly cannot be shown in figure \ref{EoS} even for a unitary increment in temperature; at the top of the domain where we will consider $T_t=-20$, we will have $\rho=-5120$.

The system described by equations \eqref{a3} and \eqref{aS} is completely defined by $\Pr$, $\Ra$, $\s$  and the dimensionless top temperature $T_t$ (recall $T_b=1$). We will call $\s$ the stiffness parameter as it is related to the buoyancy resistance of the upper fluid to overshooting convective plumes. Defining the buoyancy frequency as $N=\sqrt{-\Pr\Ra\p_z\rho}$ (based on our notations), we can see that $\s$ is also related to the ratio of the buoyancy frequency  in the stable and unstable fluid through $\s=(N^2/\p_z T)|_{z_s}/(N^2/\p_z T)|_{z_c}$, where $\cdot|_{z_s,z_c}$ means evaluated at points $z=z_s$ or $z=z_c$ in the stable or convective zone, respectively. The conductive state has a linear temperature profile $T=1-z(1-T_t)$. In this state, the fluid is convectively unstable below $z=1/(1-T_t)$, and stably stratified above, and the interface separating the unstable and stable regions corresponds to the $T=0$ (inversion temperature) isotherm. We will also consider the neutral buoyancy level $\Z$, the height at which fluid parcels rising adiabatically and without inertia from the bottom boundary would reach equilibrium.

In the next sections we solve equations \eqref{a3} with $\rho$ given by \eqref{aS} using the pseudo-spectral code Dedalus \footnote{Dedalus is available at http://dedalus-project.org.} \cite[see][and further details in appendix]{Burns2017}. We first briefly discuss the transient evolution (\S\ref{sectransients}), and then we explore the dynamical regimes and statistical properties of the coupled convective-stably stratified system at thermal equilibrium (\S\ref{sec3}-\ref{sec4}) for different stiffness parameters, $\s=2^i$ with $i\in[-8,8]$ an integer, and three different reference Rayleigh numbers, i.e. $\Ra=8\times 10^6,8\times 10^7,8\times 10^8$ (details of all 51 simulations are provided in the appendix). For simplicity, the Prandtl number is fixed to $\Pr=1$, and the top temperature is fixed to $T_t=-20$ in all simulations. The choice of $T_t=-20$ is rather arbitrary but serves the purpose of having stable and convective layers of similar height for the prototypal case shown in figure \ref{schematic}b ($\Ra=8\times 10^7$, $\s=1$), thus minimizing confinement effects on the convection and on wave propagation in the stably-stratified region. The box width relative to the height is fixed to 2, such that the effective aspect ratio for the convection is approximately 4, which limits horizontal confinement effects.

%%%%%%%%%%%%%%%
\section{Transient evolution and the height of the convection zone}\label{sectransients}
%%%%%%%%%%%%%%%

\begin{figure}
\centering
%\hspace{-0in}
\includegraphics[width=\textwidth]{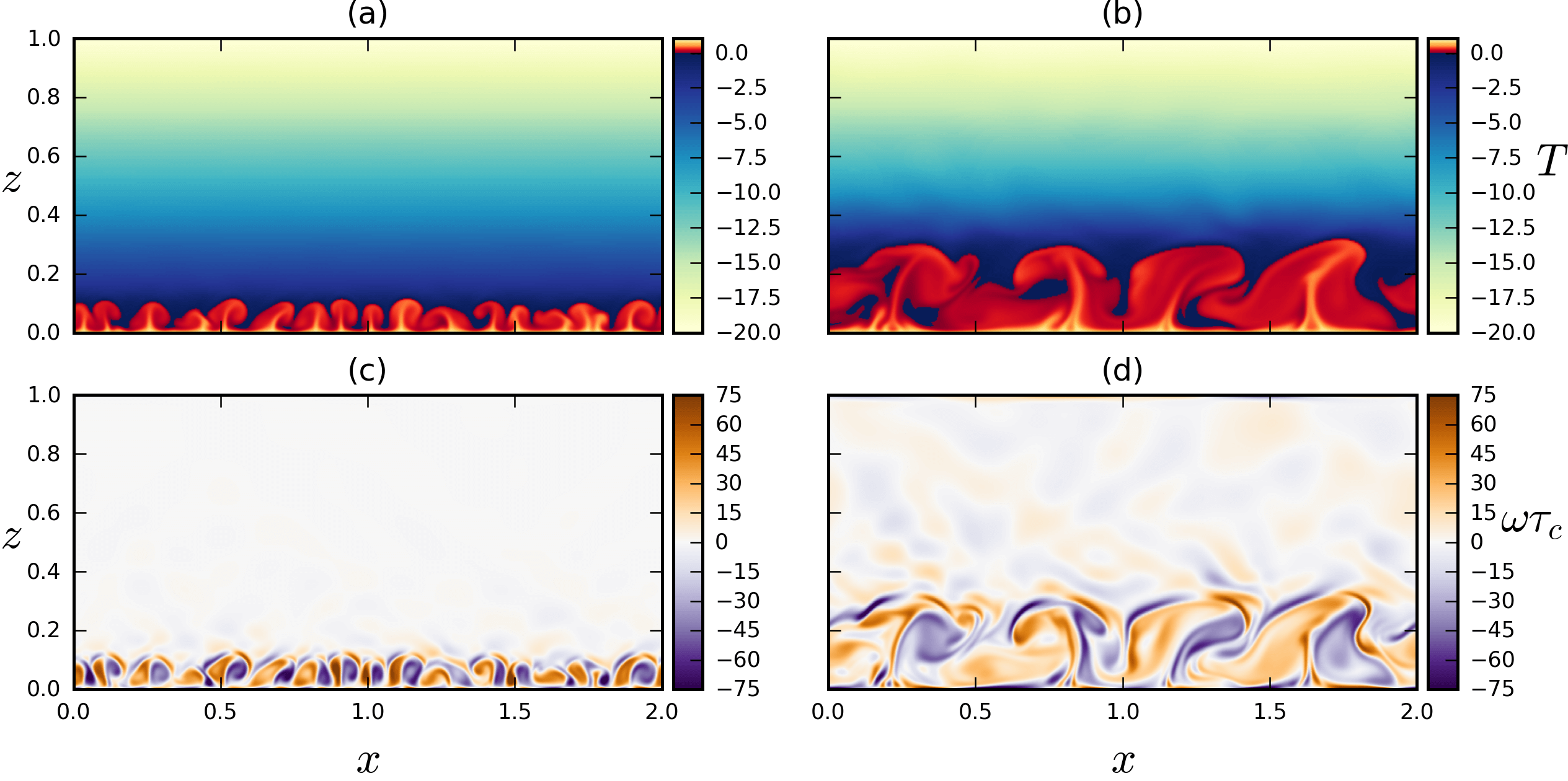}
%\vspace{-0.1in}
\caption{Transient evolution of (a)-(b) the temperature field, and (c)-(d) the normalized vorticity field $\omega \tau_c$  at times (left) $t/\tau_c=2.1$ and (right) $t/\tau_c=50.5$. The simulation is started from the static equilibrium of linear temperature profile, and the physical parameters are $\Pr=1$, $\Ra=8\times 10^7$, $\s=1$, and $T_t=-20$. $\tau_c$ is the reference turnover time scale (see text).}\label{transients}
\end{figure}

We first show in figure \ref{transients} the transient evolution of the temperature and vorticity fields between $t/\tau_c=2.1$ and $t/\tau_c=50.5$  starting from the conductive state $T=1-z(1-T_t)$ with white noise at $t=0$ for physical parameters $\Ra=8\times 10^7$ and $\s=1$ (we recall that we set $\Pr=1$ and $T_t=-20$ for all simulations in this paper); $\tau_c=1/f_c=2\pi/\sqrt{\Ra\Pr}$ is the reference turnover time scale for buoyancy forces, which will be used throughout (along with the turnover frequency $f_c$) to normalize time variables and vorticity (note that it can be obtained from equations \eqref{a3} assuming small perturbations, upon substitution of $\rho$ by \eqref{aS}, and assuming a unitary temperature gradient). After roughly two turnover times (figures \ref{transients}a,c), the convective instability develops close to the bottom boundary layer; the convective cells then expand over time upward into the overlaying stable layer (figures \ref{transients}b,d). The convective region expands because the convective flux in the convection zone is larger than the diffuse flux in the stable zone. When the convection expands, the convective flux stays about constant, but the diffusive heat flux in the stable region increases because the temperature gradient increases. The depth of the convection zone reaches a constant mean value when the heat flux through the convection zone is on average equal to the heat flux through the stable layer, which is known as the state of thermal equilibrium. 

The characteristics of the convection vary with the height of the convective region. Determining the height of the convective region at thermal equilibrium is thus a major goal of this paper. The Rayleigh number defined by \eqref{a44} is a \textit{reference} Rayleigh number, but is not a strict analogue of the Rayleigh number in classical RBC because the height of the convective region is an output of the simulation, not an input. Indeed, in our definition of the Rayleigh number we use the temperature difference $T_b^*-T_i^*$ and thermal expansion coefficient $\alpha^*_c$ related to the convection, but choose the total domain depth $\ell$ as length scale. The inversion depth $\ell/(1-T_t)$ of the conductive state could be used alternatively, but from figure \ref{transients} and previous studies on convection in water close to its density maximum, we know that the convective layer depth grows with time from the static state, and that it reaches a mean height at dynamical equilibrium that can substantially vary with the problem parameters \cite[see e.g.][and figure \ref{transients}]{Large2014}. As a result, neither $\ell$ nor $\ell/(1-T_t)$ can be expected to provide an accurate measure of the convective height. Providing a prediction for the convective layer depth including coupling between the two layers is the topic of section \S\ref{sec46}.

In a previous analysis of convection in water close to its density maximum, \cite{Moore1973} showed that a prediction of the (dimensionless) convective height $h$ could be made under several assumptions. Assuming no dynamical coupling between the stratified upper layer and the convection, the method equates the diffusive heat flux in the upper layer (approximated as $-(T_t-T_i)/(1-h)$) with the heat flux in the convection zone of the form $C\Ra_{eff}^{\beta} (T_b-T_i)/h$, where $\Ra_{eff}$ is the Rayleigh number based on the effective convective depth. The form of the convective heat flux is suggested by classical RBC studies which show that for fixed $\Pr$ the Nusselt number, $\Nu$, can be approximated as $\Nu=C \Ra^{\beta}$ with $C$ and $\beta$ two constants. Solving for the convective depth $h$ with $\Ra_{eff}=\Ra h^3$ then yields (with $T_i=0$ and $T_b=1$)
\ba{}\label{a55b}
1-h = -T_t\f{h^{1-3\beta}}{C\Ra^{\beta}},
\ea
which, under the assumption that $1-3\beta \ll 1$ (an acceptable approximation  since $0.28 <\beta < 0.31$ for laminar thermal boundary layers, cf. \cite{Grossmann2000,Ahlers2009}), reduces to
\ba{}\label{a55}
1-h \approx \f{-T_t}{C\Ra^{\beta}}.
\ea
The approximate expression \eqref{a55} for $1-h$ predicts that the convection height increases with the reference Rayleigh number $\Ra$ but decreases with $T_t$. Both effects are expected since as $T_t$ decreases, the stably stratified layer grows and the unstable layer correspondingly shrinks in the conductive state. The prediction, however, is independent of the stratification parameter $\s$, as the dynamical coupling between the stably stratified and convective regions are neglected. We will show that this simplification is appropriate only in the large stiffness regime ($\s \gg 1$), and that the convection height cannot be in general inferred from this model.

%%%%%%%%%%%%%%%
\section{Dynamics at thermal equilibrium for different stiffnesses}\label{sec3}
%%%%%%%%%%%%%%%

We now turn our attention to the system dynamics at thermal equilibrium, i.e. once statistical steady-state and depth-invariant heat flux are achieved. Unlike in figure \ref{transients}, in order to avoid the thermal equilibration time for the stable layer, we start our simulations from a temperature profile close to what is expected at thermal equibrium (see appendix). 

We show in figure \ref{schematic}a-c three snapshots obtained at thermal equilibrium for (a) high $\s=2^8$, (b) moderate $\s=1$, and (c) small $\s=2^{-8}$ stiffness. We select snapshots for  $\Ra=8\times 10^7$, since this intermediate Rayleigh number case will be the focus of the detailed analysis in section \S\ref{sec4} ($\Pr=1$, $T_t=-20$). In figures \ref{schematic}a,b, the density anomaly $\rho$ is shown from the bottom up to the instantaneous neutral buoyancy level (i.e. $z=\Z(x)$, where $\rho(x,\Z)=-1$, labeled as the \textit{interface} in figures \ref{schematic}), while the vorticity is shown above it. The idea of showing both density and vorticity such as in figure \ref{schematic} is not new \cite{Lecoanet2015}, and allows to combine in a single plot the information on buoyancy effects in the convection zone and waves in the stable layer. In figure \ref{schematic}c, the density anomaly is shown everywhere because plumes can in theory rise all the way to the top of the domain; the stratification is indeed so weak that the lightest fluid is within the bottom boundary layer.

Figure \ref{schematic}a, obtained for a high stiffness parameter, shows two distinct regions above and below $\Z$, which coincides with the $T_i$ isotherm. The bottom layer consists of RB convective cells made of up-down symmetric plumes rising between the bottom plate and the neutral buoyancy level. The top layer is stably stratified and supports waves. The interface is almost perfectly straight and we expect that the waves are excited by Reynolds stress forcing \cite[][]{Lecoanet2015}. Global internal gravity modes  are present, with approximately three wavelengths fitting along the vertical direction.

\begin{figure}
\centering
%\hspace{-0in}
\includegraphics[width=0.9\textwidth]{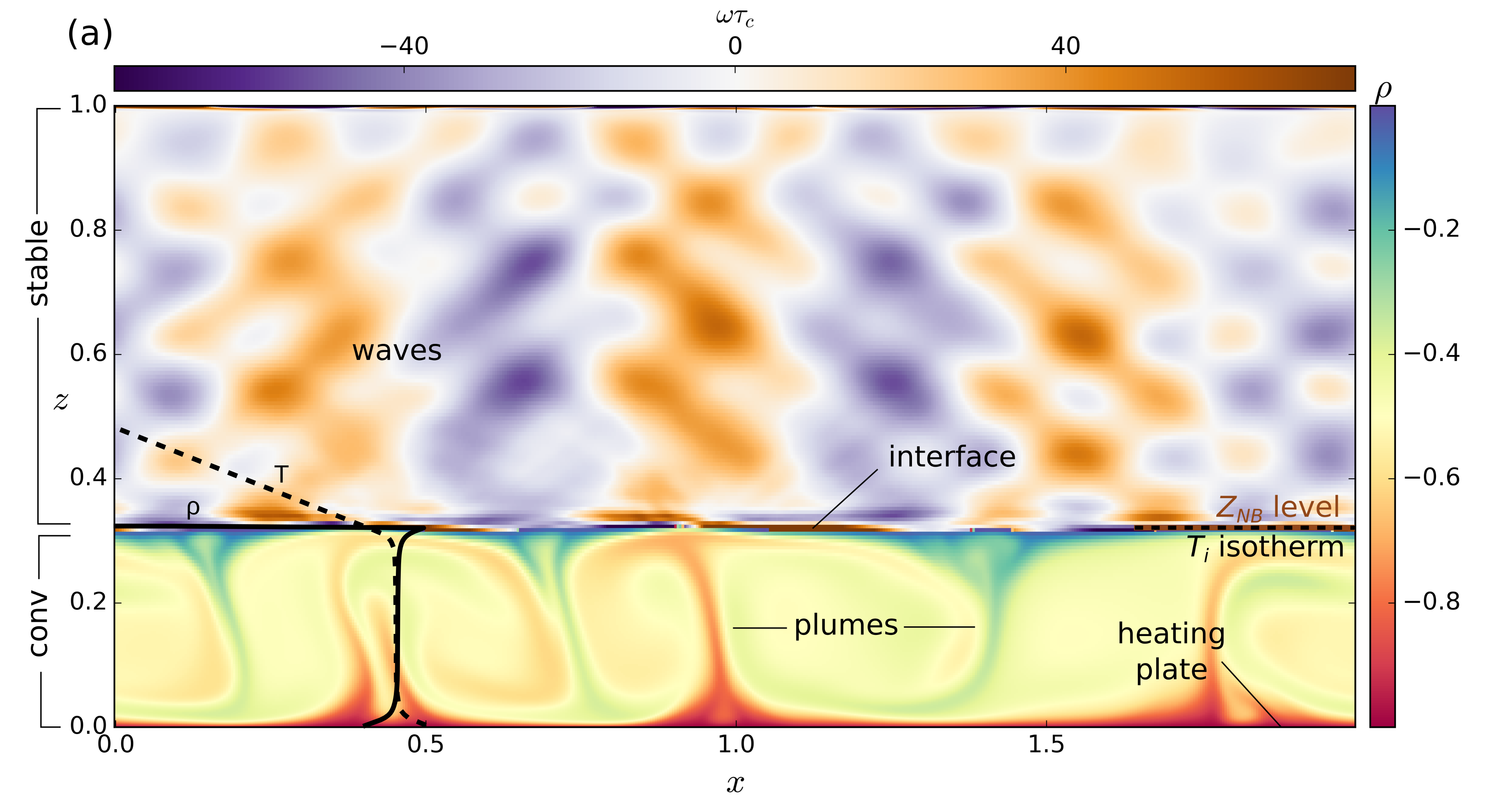}
\includegraphics[width=0.9\textwidth]{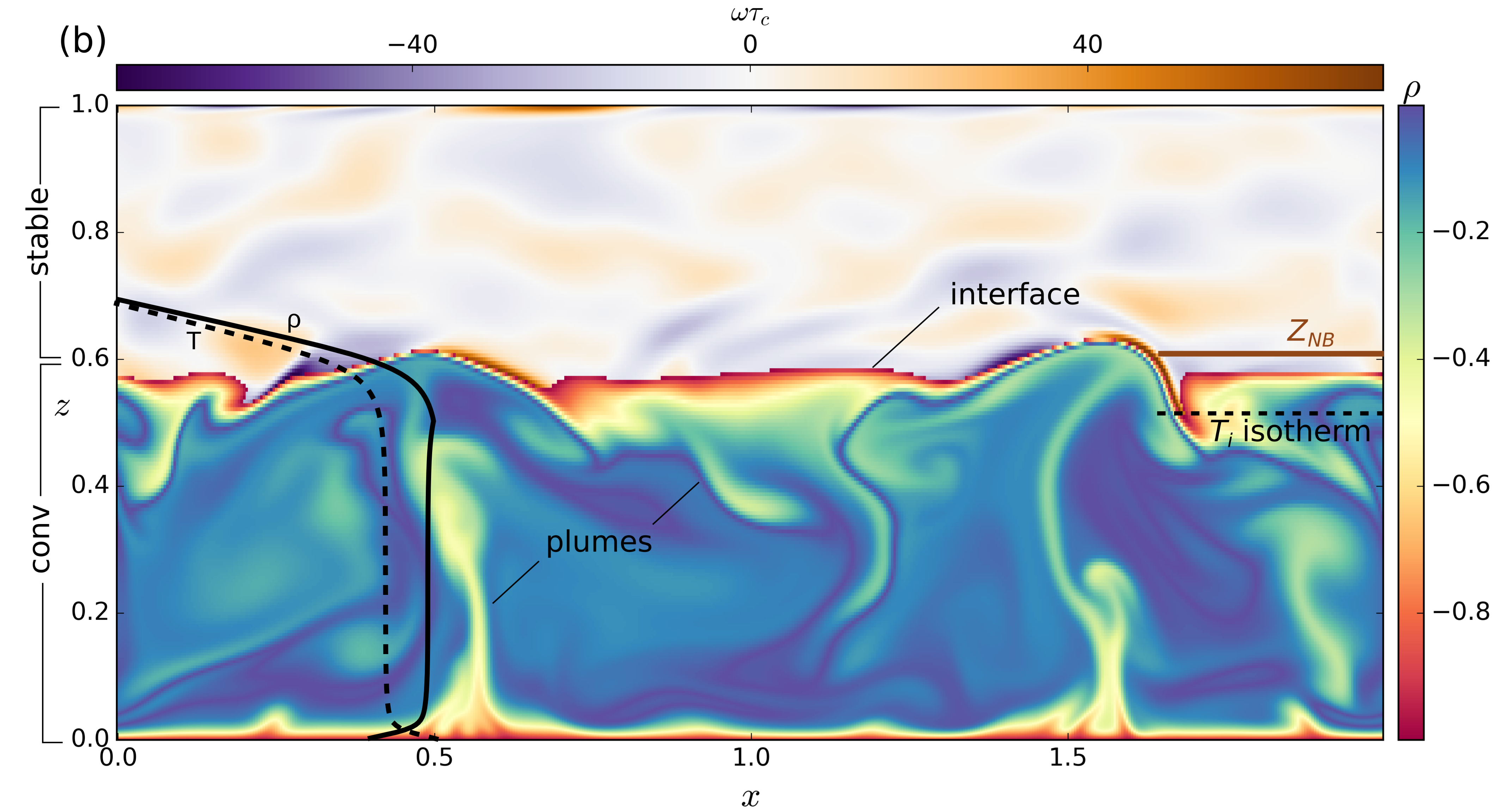}
\caption{See caption on next figure.}
\end{figure} 

\begin{figure}
\ContinuedFloat
\centering
%\hspace{-0in}
\includegraphics[width=0.9\textwidth]{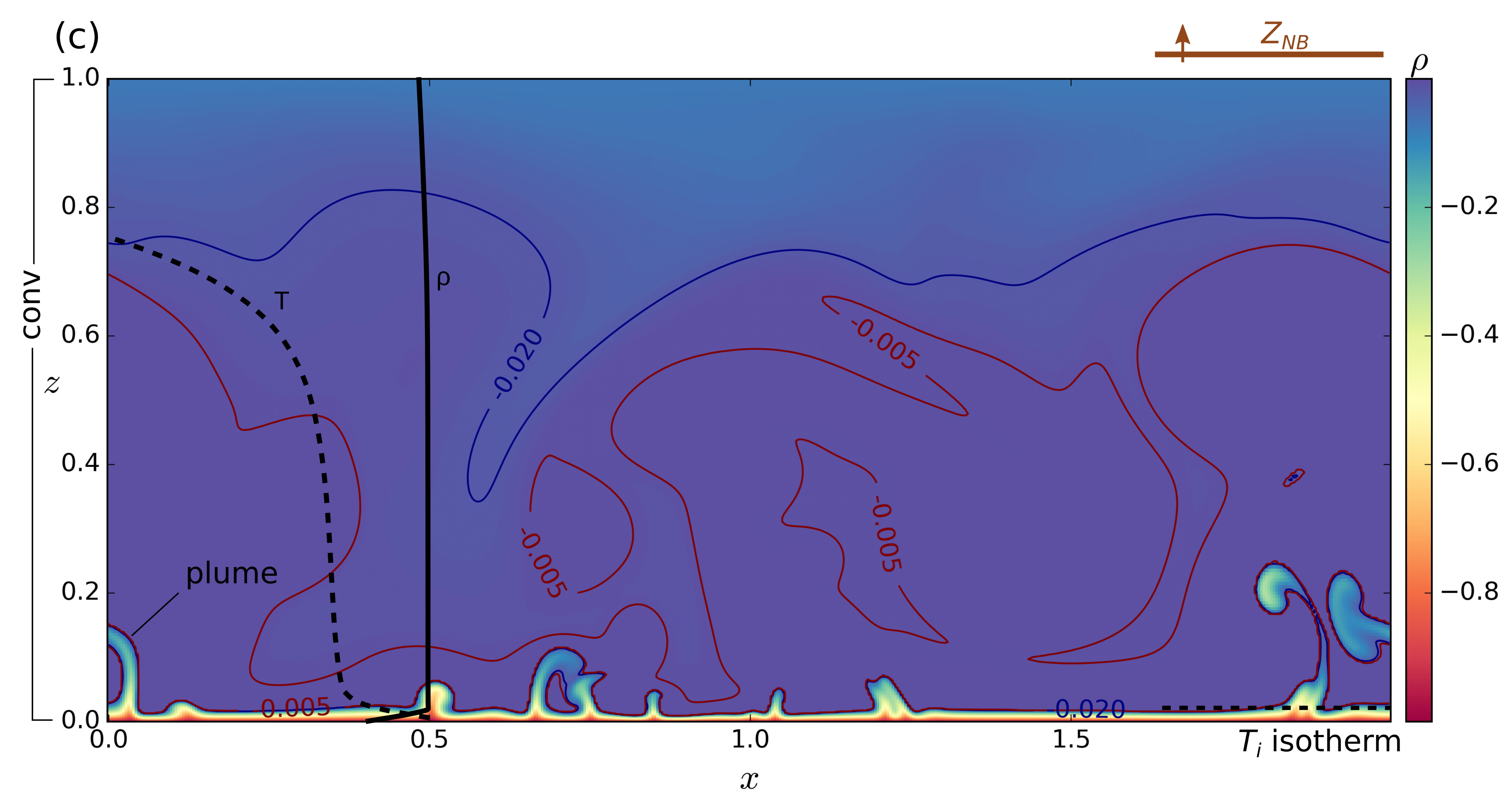}
%\vspace{-0.1in}
\caption{(Cont.) Simulation snapshots for $\Ra=8\times 10^7$ and stiffness (a) $\s=2^8$, (b) $\s=1$, (c) $\s=2^{-8}$ representatives of the system's dynamic at thermal equilibrium, i.e. obtained after several thousands convective turnover times. The density anomaly is shown in the convection zone, below the neutral buoyancy height $\Z(x)$ (i.e. where $\rho(x,\Z)=-1$). Above $\Z$, the fluid is stably stratified, and we plot the vorticity. In (a) convective cells with a top/down symmetry are seen in the lower layer, generating global wave modes in the upper stably-stratified layer. In (b), the convection zone is larger but significant coupling between the two layers is obtained as shown by the strong variations of $\Z$ with $x$. In (c) the system changes significantly with rising plumes close to the bottom interacting with large-scale structures (shown by contour lines) higher in the fluid (note that $\Z$ is outside of the domain such that we show the density field everywhere in this case). We also show the mean temperature and anomaly density profiles by the dashed and solid lines as later discussed and reported in figure \ref{mean}. Movies of figures  \ref{transients} and  \ref{schematic} can be found online at sites.google.com/site/fludyco  in the outreach section.}\label{schematic}
\end{figure}

Figure \ref{schematic}b, obtained for stiffness $\s=1$, exhibits the same features as figure \ref{schematic}a. Plumes are emitted from the bottom plate and the \textit{interface}, while internal waves are observed above $\Z$. The aspect ratio of the convective cells is roughly 1, as in figure \ref{schematic}a, but because the interface lies higher, only 2 cells are obtained, compared to 3 for $\s=2^8$. Compared to the high-stiffness case, however, the instantaneous interface is much more distorted as a result of upward-going plumes impinging on the relatively weak stable stratification. Rising and sinking plumes do not have the same symmetry either: down-going heavy plumes look like filaments, entraining chunks of lighter fluid with them. These observations are also reflected in the difference of location of the $T_i$ isotherm and neutral buoyancy level, which can be seen as a first approximation of the overshooting length scale of strong penetrative convection. Note that this overshooting length scale is defined from the mixing of stable fluid into the convective region (sinking of $T_i$ isotherm within the convection zone), and is therefore not exactly equivalent to the overshooting length scale  based on the mixing of convective fluid into the stable region considered in penetrative convection studies on top of a background state. The more chaotic dynamical transition between the convective and stably-stratified regions clearly results in a less organized internal gravity wavefield than in the high-stiffness case of figure \ref{schematic}a.

\begin{figure}
\centering
%\hspace{-0in}
\includegraphics[width=0.85\textwidth]{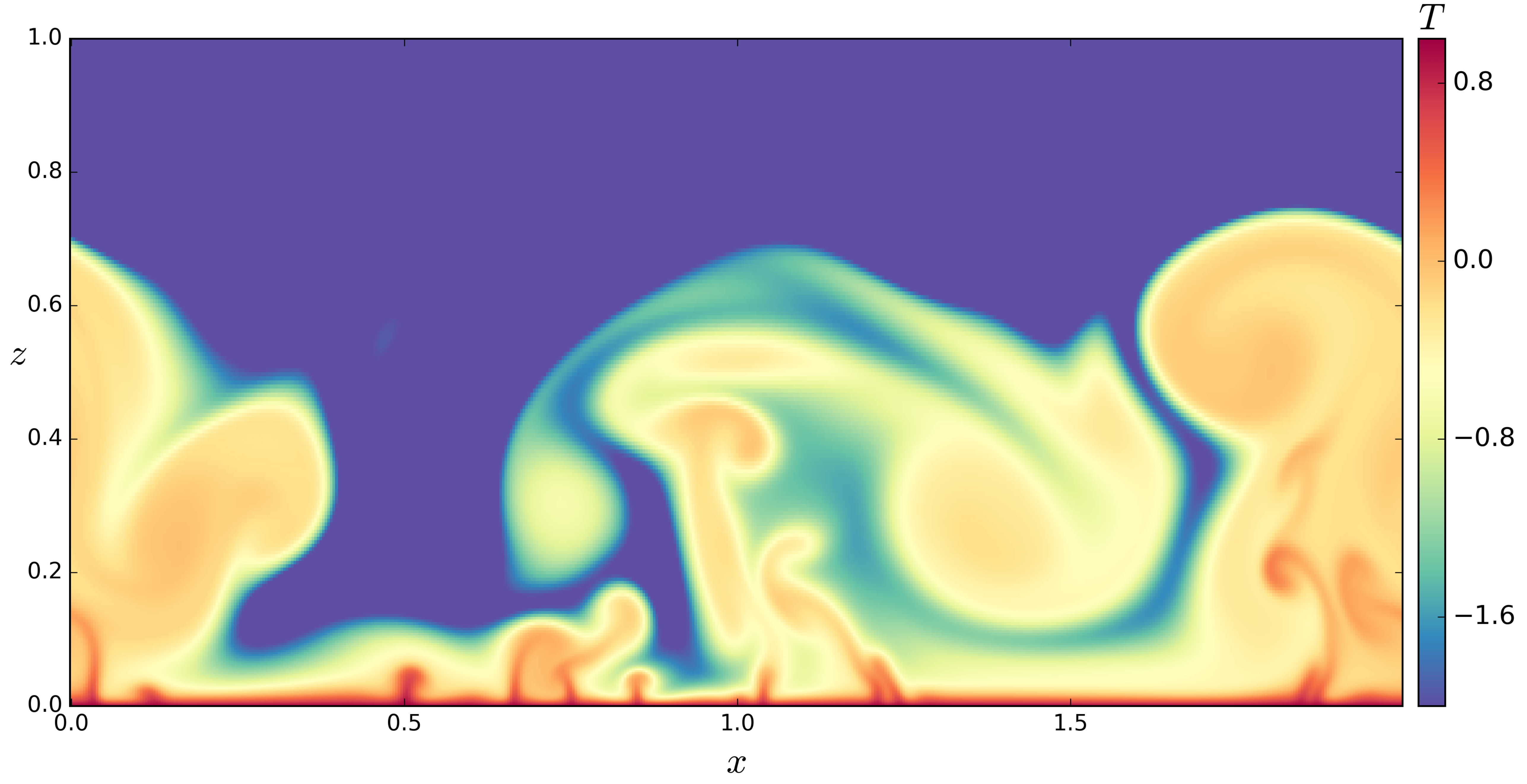}
%\vspace{-0.1in}
\caption{Snapshot of the temperature field corresponding to the density anomaly plot of figure 
\ref{schematic}c.}\label{temp}
\end{figure}

Figure \ref{schematic}c, obtained for the smallest stiffness parameter $\s=2^{-8}$ considered, shows a very different behavior than the previous two cases. In this simulation, the bulk of the fluid is everywhere heavier than near the bottom plate. A stark difference also comes from the fact that there are only upward-going plumes.  This is because the $T_i$ isotherm is within the thermal bottom boundary layer, so convection only occurs near $z=0$. The sinking of the $T_i$ isotherm is an important result that can be explained from the fact that the top stable layer, which is initially quiescent in the simulations, is so heavy for small $\s$ that it can squeeze the buoyancy-driven convective layer down as soon as it is destabilized by overshooting convective plumes. Because the stable layer is only marginally stratified for small $\s$, the entire stable fluid is put into motion and slumps down when the bottom part is destabilized by the plumes. This explains why entrained fluid motions  extend  almost everywhere in the fluid. Note that the cold entrained fluid moves seemingly passively, i.e. slower than the plumes, but because of its inertia and large temperature anomaly still dominates dynamically almost everywhere and brings the bulk density anomaly close to zero. Looking at the temperature field at the same simulation time (cf. figure \ref{temp}) shows that hot rising  plumes still impact high parts of the fluid as a result of their inertia.  Thus, plumes play an essential role in the dynamics by maintaining the entrained fluid in motion, but not necessarily in the heat transfer. As they rise further up, plumes broaden (thereby carving the large-scale entrained fluid visible in figure \ref{schematic}c) and cool down to the point that they can contribute negatively to the (positive) heat transfer. Nonetheless, as will be demonstrated later, convection is efficient from $z=0$ up to $z = 0.9$, because the large-scale entrained structures of low density anomaly (in part visible from contour lines of iso-density in figure \ref{schematic}c) are efficient at transporting cold fluid down. In the weakly-stratified limit, the overshooting length scale is of the same order as the entire fluid depth. Note that the top of the fluid layer is stably-stratified, although weakly.

%%%%%%%%%%%%%%%%%%
\section{Statistical analysis and regime properties}\label{sec4}
%%%%%%%%%%%%%%%%%%

The simulation snapshots of figure \ref{schematic} show that the stiffness parameter $\s$ plays a key role in the coupling between the convective and stably-stratified regions. The high-stiffness regime shows strong top-down symmetry in the convection zone, suggesting negligible feedback from the waves on the convection. This symmetry is no longer present for moderate stiffness $\s\sim 1$, and the dynamics completely change for low stiffness.

We now quantitatively study the effect of the stiffness $\s$ on the interaction between the convective and stably stratified regions and on the heat transfer. The results are obtained for simulations in thermal equilibrium, i.e. such that the statistics are temporally converged and the total heat flux is depth-invariant. Reaching thermal equilibrium requires running simulations over several dissipation times, which requires long run times. Use of judicious initial conditions allows a more rapid convergence to thermal equilibrium (see appendix for more details). We will make extensive use of the $x$ average operator $\langle ~\cdot~ \rangle_x = \int dx/2$, ($2$ is the box width); the volume average operator $\langle ~\cdot~ \rangle = \int dz dx/2$; and the time average operator, $\overline{~\cdot~}$. Time averages are typically performed over tens to thousands turnover times depending on $\s$ (see appendix for details).

\begin{figure}
\centering
%\hspace{-0in}
\includegraphics[width=0.8\textwidth]{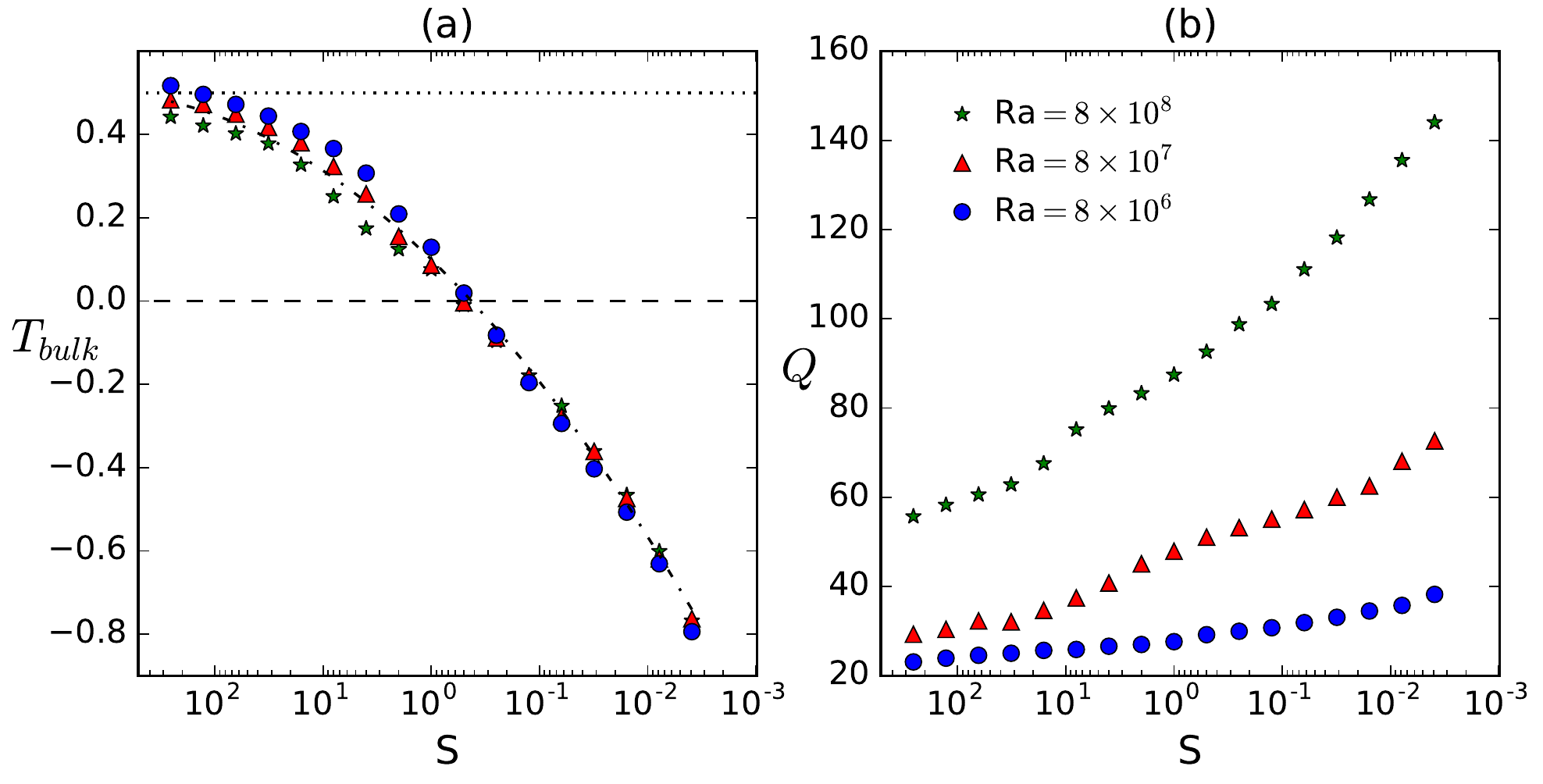}
\vspace{-.0in}
\caption{(a) Bulk temperature $T_{bulk}$ as a function of $\s$ ($\s$ decreases to the right) averaged over the convection zone (see text for details). The dash-dot line is a best fit obtained for all data points and given by equation \eqref{tfit}. The dashed line highlights the inversion temperature $T_i=0$ where $\alpha$ changes sign, and the dotted line represents the bulk temperature expected from classical Rayleigh-B\' enard convection. (b) Total heat flux $Q$ as a function of $\s$. Results are shown for three different reference Rayleigh numbers $\Ra$. }\label{heat}
\end{figure}

We first present in figure \ref{heat} simulation results obtained for 3 different Rayleigh numbers ($\Ra=8\times 10^6,8\times 10^7,8\times 10^8$) and 17 different stiffness parameters ($\s$ from $2^8$ to $2^{-8}$; see table \ref{table} in appendix). In figure \ref{heat}a we show the bulk temperature, i.e. $T_{bulk}=\int_z dz\mt$, averaged in $z$ over the convection zone (i.e. where the convective heat transfer $q_c$ accounts for at least 95\% of the total heat flux $q$). For all three Rayleigh numbers we can see that the bulk temperature increases with the stiffness parameter. In the limit of large stiffness, $T_{bulk}\rightarrow 0.5$, which is the temperature expected from a convection-only simulation with top/bottom temperatures of $T_t=0$ and $T_b=1$. Thus the large-stiffness limit of the bulk temperature tends to the classical RBC bulk temperature even though the EoS is nonlinear. This is consistent with figure \ref{schematic}a, where the density anomaly in the bulk is roughly $\rho=-0.5$, corresponding to $T_{bulk}=0.5$. In the small-stiffness limit, the bulk temperature decreases below the $T_i=0$ isotherm, suggesting that significant mixing of low-temperature fluid occurs within the convective region. The overlap of the results for all $\Ra$ further indicates that this mixing is relatively independent of the degree of turbulence. As long as the lower layer is convectively unstable and plumes are emitted, they can destabilize the upper stable layer entirely, because it is only marginally stratified. For high stiffness, higher $\Ra$ tends to increase the entrainment/mixing and lower the bulk temperature. The data for all $\Ra$ are well approximated in the range $\s\in[2^{-8},2^8]$ by a best polynomial fit law in terms of $\ln(\s)$, i.e.
\ba{}\label{tfit}
T_{bulk}(\s)=-0.0075\ln(\s)^2+0.11\ln(\s)+0.1,
\ea
shown as the dash-dot line in figure \ref{heat}a. This empirical law will be used in \S\ref{sec46} to provide an estimate for the depth of the convection zone. 

We define the $z$-dependent convective and diffusive heat fluxes to be
\bsa{}
q_c &=\overline{\langle wT \rangle}_x, \\
q_d &=\overline{\langle -\p_zT \rangle}_x
\esa
and we recall that the total heat flux $q=q_c+q_d$ can be shown to be independent of $z$ when averaged over long time periods (recall that time and therefore $w$ are normalized by the thermal time, and that $\Pr=1$). Figure \ref{heat}b then shows the total heat transfer $Q=\int q dz$ as a function of the stiffness parameter ($z$-averaging removes the small $2\%$ maximum relative discrepancy obtained using the infinity norm in our simulations, showing the good statistical convergence). Because we work with dimensionless variables, we note that $Q$ is normalized by a diffusive heat flux, and thus may be interpreted as a Nusselt number. We refrain from interpreting $Q$ as a Nusselt number, however, because the Nusselt number definition requires a choice of length scale (see section \S\ref{sec45}). The heat transfer $Q$  increases with decreasing stiffness $\s$ for all three Rayleigh numbers, which is in agreement with the observation that the diffusive heat flux in the stable layer should increase as a result of the growing convection zone (see figure \ref{schematic}). As expected, we also observe that $Q$ increases with $\Ra$ for all $\s$. In the high-stiffness limit, the slope $|dQ/dS|\rightarrow 0$, which suggests that the classical RBC regime might be recovered for large $\s$. This further implies that the depth of the convection zone $h$ becomes independent of $\s$ since the heat flux in the stable layer is dominated by diffusion, so $Q \approx -T_t/(1-h)$. However, how large $\s$ should be to recover RBC clearly depends on $\Ra$, since for the highest $\Ra$ simulations, $Q$ still changes appreciably for $\s=256$.  For small $\s$, on the other hand, $Q$ increases continuously with decreasing $\s$. Thus,  different mechanisms occur when varying $\s$, as could be inferred from figures \ref{schematic} and \ref{temp}.

%%%%%%%%
\subsection{Vertical profiles of the mean temperature and density}\label{sec41}
%%%%%%%%

\begin{figure}
\centering
%\hspace{-1.2in}
\includegraphics[width=0.8\textwidth]{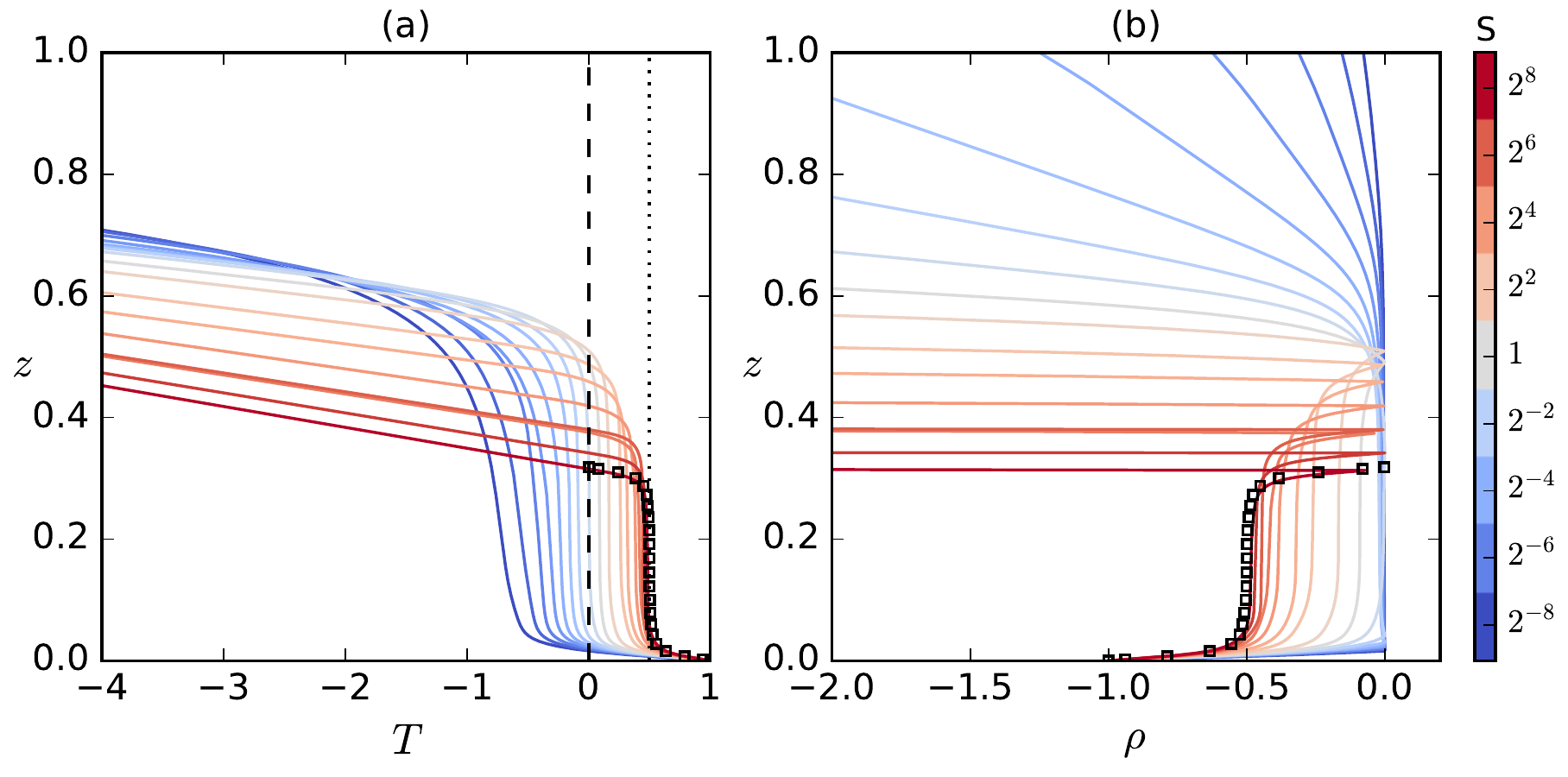}
\vspace{-.0in}
\caption{(a) Mean temperature $\mt$ and (b) mean density anomaly $\overline{\langle \rho \rangle}_x$ profiles as functions of $z$ for $\Ra=8\times 10^7$ and for different stiffness parameters $\s$. For strong stratification, the density anomaly in the convection zone is constant within a bulk bounded by two thin layers where the density quickly increases, reminiscent of classical RBC. For weak stratification, the fluid density only increases within the bottom boundary layer and is almost uniform or weakly decreasing above. As in figure \ref{heat}, the vertical dashed line in (a) highlights the inversion temperature $T_i=0$, and the dotted line indicates the bulk temperature expected from classical Rayleigh-B\' enard convection. The unfilled-square symbols show results from a convection-only simulation with rigid no-slip top/bottom boundaries and a vertical height $h=0.32$, which is an estimate of the depth of the convection zone for $\s=2^8$ (obtained from $h=(2L_{db}+L_{cp}+L_{ce})$, see \eqref{a13}).}\label{mean}
\end{figure}

In figure \ref{mean}, the mean temperature $\mt$ and density $\mr$ profiles are shown as a function of $z$. Because the convection is steady for $\Ra=8\times 10^6$ at high $\s$, and vertical confinement effects may affect the higher $\Ra=8\times 10^8$ simulations (due to the higher convective flux resulting in a larger $h$, and to overshoot), from here onward we choose to focus on the case $\Ra=8\times 10^7$ (obtaining converged statistics would also require a much longer time for $\Ra=8\times 10^8$ than for $\Ra=8\times 10^7$). The trends regarding the effect of $\s$ are, nevertheless, similar for all three Rayleigh numbers considered. We use colors ranging from blue to red to represent increasing stiffness. In figure \ref{mean}a, we plot the temperature between $-4$ and $1$ because the temperature decreases linearly with $z$ for $\mt\leq -4$ (recall $T_{t}=-20$). The highest-stiffness case (dark red curve) shows a symmetric temperature profile with $z$ for $\mt\in[0,~1]$, i.e. between the bottom and inversion temperature, similar to what is obtained from a purely convection simulation (see unfilled-square symbols). The mean temperature in the well-mixed region is roughly 0.5, as observed in figure \ref{heat}a. As the stiffness decreases, the mean temperature profiles become less symmetric; eventually  the bulk temperature drops below 0. It is relatively surprising that for $\s \leq 2^{-2}$ the $T_i=0$ isotherm is found within the thermal bottom boundary layer, so the fluid is on average stably-stratified from the outside of the bottom boundary layer all the way to the top of the domain. Nonetheless, it is in agreement with the observations made on figure \ref{schematic}c, i.e. that entrained cold fluid motions squeeze the buoyancy-driven convection zone close to the bottom boundary layer.

The stable stratification of the entire fluid domain, except the bottom boundary layer, for low stiffness $\s$ can be clearly seen in the (blueish) $\mr$ profiles of figure \ref{mean}b. For $\s \leq 2^{-2}$ the density increases rapidly from -1 at $z=0$ to 0, and then becomes roughly uniform or slowly decreasing with $z$ (as observed in figure \ref{schematic}c). For high stiffness, on the other hand, large positive density gradients are obtained in two distinct regions separated by the bulk region, such that density increases over a relatively large vertical extent. It should be noted that because density anomaly gradients vary strongly where the dynamics change from stable to unstable, the buoyancy frequency $N$ varies strongly there too. However, since we chose a constant background density and an EoS for which $\alpha$ is constant in the stable layer, $N$ is mostly constant deep in the stable region. Had we chosen a non-uniform background density or an EoS for which $N$ would be variable deep in the stable region on length scales large in comparison to the vertical wavelength of the wave, then wave propagation would follow WKB ray theory \cite{dosser2011}.

Unstable (positive) density gradients are only found close to the bottom boundary for low stiffness $\s$ because buoyancy-driven convection is limited to a thin bottom layer when the stratification is weak. Nonetheless, it is clear from figure \ref{mean}b that the vertical extent over which the density is approximately uniform always increases with decreasing stiffness $\s$, such that there might be a mixed region of non-negligible extent for all values of $\s$ considered. We will confirm that convective motions take place over a volume fraction that grows monotonously with decreasing stiffness in the next section.

%%%%%%%%
\subsection{Heat transfer and entrainment}\label{sec42}
%%%%%%%%

\begin{figure}
\centering
%\hspace{-1.2in}
\includegraphics[width=0.95\textwidth]{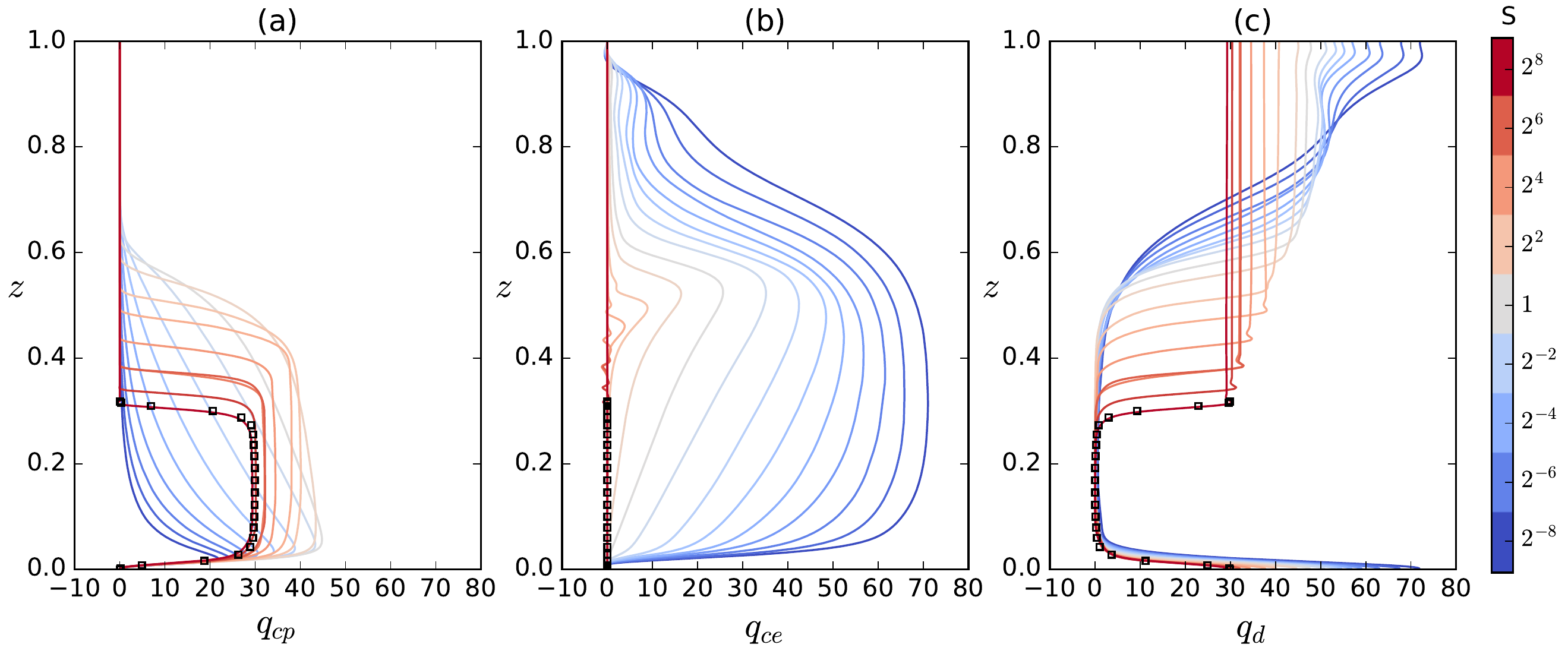}
\vspace{-.0in}
\caption{Vertical profiles of the heat transfer due to (a) convective plumes $q_{cp}$, (b) entrainment $q_{ce}$, and (c) diffusion $q_d$ for $\Ra=8\times 10^7$. Different curves correspond to different stiffness $\s$. For high stiffness, we obtain a symmetry in the heat transfer in the convection zone due to plumes ($q_{cp}$) in (a), suggesting classical Rayleigh-B\' enard convection. The contribution from entrainment to the total heat transfer can be seen in (b) to be dominant in the low-stiffness cases and negligible in the high-stiffness regime. The total heat transfer  $q=q_{cp}+q_{ce}+q_d$ is constant along the $z$ axis up to 2\% maximum relative discrepancy. As in figure \ref{mean}, the unfilled-square symbols show results from a convection-only simulation with a vertical height comparable to the convection depth of the highest-stiffness case.}\label{qc}
\end{figure}

In order to explain the increase of $Q$ with decreasing $\s$ from figure \ref{heat}b, as well as the--first increase, then decrease--of the layer depth over which buoyancy-driven convection occurs  (figure \ref{mean}), we split the convective heat transfer $q_c$ into two contributions, i.e. 
\bsa{qqq}
q_{cp}=\overline{\langle wT\mathcal{H}\lb +(T-T_i)\rb \rangle}_x, \\
q_{ce}=\overline{\langle wT\mathcal{H}\lb -(T-T_i)\rb \rangle}_x,
\esa
where $\mathcal{H}$ is the Heaviside step function (recall $T_i=0$). The heat transfer $q_{cp}$ gives the contribution from convective motions occurring as a result of buoyancy effects, and is equivalent to the convective heat flux in classical Rayleigh-B\' enard convection. $q_{ce}$ measures the contribution to the heat flux from fluid that is entrained (hence the superscript $e$), i.e.~from fluid parcels expected to be in a stable region ($T<T_i$) and that yet have a net negative velocity. We recall that the sum  $q_{cp}+q_{ce}+q_d=q=Q$ is constant for all $z$.

From figure \ref{qc}a, we find for the high-stiffness cases ($\s \in [2^{2},2^8]$) a symmetric profile for $q_{cp}$ in the convection zone, with a rapid increase close to the bottom boundary and a rapid decrease higher up, where density gradients change sign (see figure \ref{mean}b). This is a typical profile of the total convective heat flux in classical RBC (see unfilled-square symbols). As $\s$ decreases from $2^8$ to $2^2$, however, $q_{cp}$ increases in parallel with increasing convective layer depth (which can be taken approximately as the height at which $q_{cp}$ decreases quickly).  This is qualitatively different from RBC, which exhibits a weak dependence of $q_{cp}$ on $h$. For $\s \leq 2^1$, $q_{cp}$ loses its symmetry and decreases as the stiffness decreases (gray to dark blue). For low stiffness, the buoyancy-driven heat flux reaches a maximum close to the bottom boundary layer, and decays rapidly with $z$ away from it.

The dominant contribution to convective heat transfers does not come from buoyancy effects for low stiffness $\s$, but from fluid motions in an overall stably-stratified environment, as can be seen from figure \ref{qc}b. The heat transfer $q_{ce}$ is almost uniformly 0 for high-stiffness cases, but is large for low stiffness $\s$; both the maximum of $q_{ce}$ and the vertical extent increase with decreasing stiffness. As $\s$ decreases from $2^8$ to approximately 1, $q_{ce}$ becomes non-negligible close to the density inversion height as a result of plumes overshooting in the stable layer and inducing some entrainment of the stable fluid $(T<T_i)$ \cite[see e.g.][]{Brummell2002}. For low stiffness simulations,  the entrained fluid has a density roughly equal to the density of the fluid just outside of the thermal boundary layer (see figure \ref{mean}b), so the entrained fluid has a strong influence on the rising thermal plumes. Buoyancy-driven effects become confined closer and closer to the bottom boundary as $\s$ decreases, but at the same time plumes can entrain more of the stably-stratified fluid since the stable buoyancy frequency decreases. This competition results in higher heat transfers for smaller stiffness parameters (as seen in \ref{heat}b), and a smaller diffusive upper layer close to the top boundary of the fluid (figure \ref{qc}c). It should be noted that the efficiency of heat transfers by the entrained fluid depends on the temperature anomaly carried by the stable fluid. Here, the top temperature is $T_t=-20$, which means that the entrained fluid can have temperature anomalies in the range $[-20,0]$. This range is broader than the typical temperature anomaly of buoyant plumes of 0.5 to 1 (see e.g. figure \ref{pdft}), hence partly explaining the relative high values of $q_{ce}$ compared to $q_{cp}$ in spite of the relatively slow velocities of the entrained fluid (see section \ref{sec44}).

\begin{figure}
%\centering
\hspace{.6in}
\includegraphics[width=0.75\textwidth]{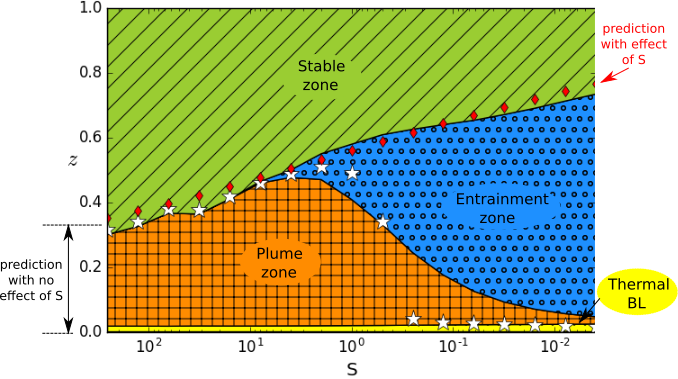}
\vspace{-.0in}
\caption{Volume fractions occupied by the bottom thermal boundary layer, the plume zone, the entrainment zone, and the stably-stratified top layer as functions of the stiffness parameter $\s$ and for $\Ra=8\times 10^7$ (see figure \ref{supp8eRa} in appendix for $\Ra=8\times 10^8$). The layer thicknesses are computed based on the profiles shown in figure \ref{qc} and according to equations \eqref{a13}. For each stiffness parameter, the position where the average temperature in the simulations is equal to the inversion temperature is shown by a star. It can be clearly seen that the inversion temperature isotherm is on average close to or within the bottom thermal boundary layer for $\s \leq 2^{-2}$. The double-headed arrow on the left-hand side shows the convection layer thickness $h$ prediction based on equation \eqref{a55b}. The filled diamonds show the empirical prediction of $h$ based on equation \eqref{hpred} presented in section \ref{sec46}.}\label{layers}
\end{figure}

The respective contributions from plumes, entrained fluid and diffusive effects to the heat transfer for varying stiffness are summarized in figure \ref{layers} for $\Ra=8\times 10^7$ (and in figure \ref{supp8eRa} in appendix for $\Ra=8\times 10^8$). Each region represents the volume fraction where either $q_{cp}$, $q_{ce}$, or $q_d$, dominates over the others, and each region thickness is computed based on the corresponding heat flux variables as
\ba{}\label{a13}
L_{cp} = \int_0^1 \f{q_{cp}}{Q} dz,\quad L_{ce} = \int_0^1 \f{q_{ce}}{Q} dz,\quad L_{db} = \int_0^{z_*} \f{q_d}{Q} \f{dz}{z_*},\quad L_{ds} = \int_{z_*}^1 \f{q_d}{Q} \f{dz}{1-z_*}.
\ea 
The thicknesses of the two diffusive regions, i.e. of the thermal bottom boundary layer and the upper stable layer, denoted by $L_{db}$ and $L_{ds}$, are computed from the same diffusive heat flux $q_d$, but we separate the integral values over the two zones at $z=z_*$ with $z_*$ taken in the middle of the mixed region (i.e. defined as $\lcb z,~q_d(z)\leq 5\% Q \rcb$), in order to avoid an overlap. The choice of $z_*$ is rather arbitrary, but results in an appropriate estimate for the depths of the bottom boundary layer and upper stable layer. Importantly, we have $L_{cp}+L_{ce}+L_{db}+L_{ds}=1$, which is the total domain height.

Figure \ref{layers} clearly shows that plumes dominate convective heat transfers for high stiffness (left-hand-side of figure \ref{layers}). For low stiffness (right-hand-side), however, it is the entrained fluid that dominates over most of the mixed region, and plume-driven heat fluxes are confined near the bottom plate. The vertical extent of the mixed region increases monotonically with decreasing $\s$, and the stratified upper layer correspondingly shrinks, which is in agreement with increasing heat transfer (cf. figure \ref{heat}). The white stars show the position of the $T_i$ isotherm in our simulations, and it can be seen that this isotherm sinks close to or within the thermal bottom boundary layer for $\s\leq 2^{-2}$. The fact that the change of the $T_i$ position is much more sudden than the change of the plume region relative to the entrainment region indicates that plumes carrying heat upward are still somewhat effective despite the low bulk temperature; this is in agreement with figure \ref{temp} in which plumes that are still relatively hot can be seen as high as $z\sim 0.6$. In the high-stiffness regime it is expected that the convective layer thickness $h$ is well approximated by equation \eqref{a55b}. This is verified in figure \ref{layers} as the predicted $h$ (shown by the double-headed arrow), including the plume region as well as two symmetric thermal boundary layers, is in good agreement with the highest-stiffness $\s=2^8$ results (note that we use $\beta=0.27$ and $C=1/5.5$ in equation \eqref{a55b} based on convection-only simulation results that we ran with Dedalus and obtained for Rayleigh numbers in the range $[10^6,10^9]$). The fact that the convection height for $\s=2^7$ is different than for $\s=2^8$, however, suggests that the stiffness must be quite large ($>\co(100)$) to be in the asymptotic high-stiffness regime.

%%%%%%%%
\subsection{Kinetic energy density spectra}\label{sec43}
%%%%%%%%

\begin{figure}
\centering
%\hspace{-1.2in}
\includegraphics[width=0.32\textwidth]{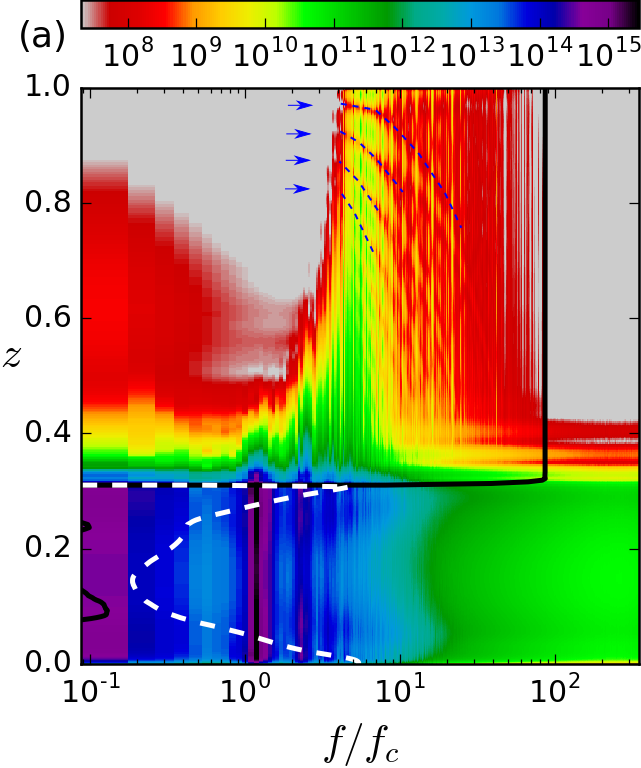}
\includegraphics[width=0.32\textwidth]{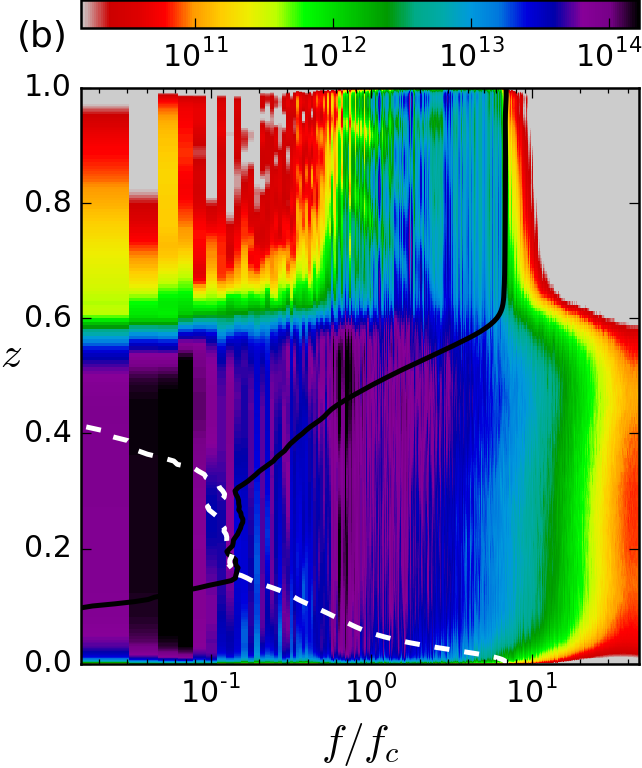}
\includegraphics[width=0.32\textwidth]{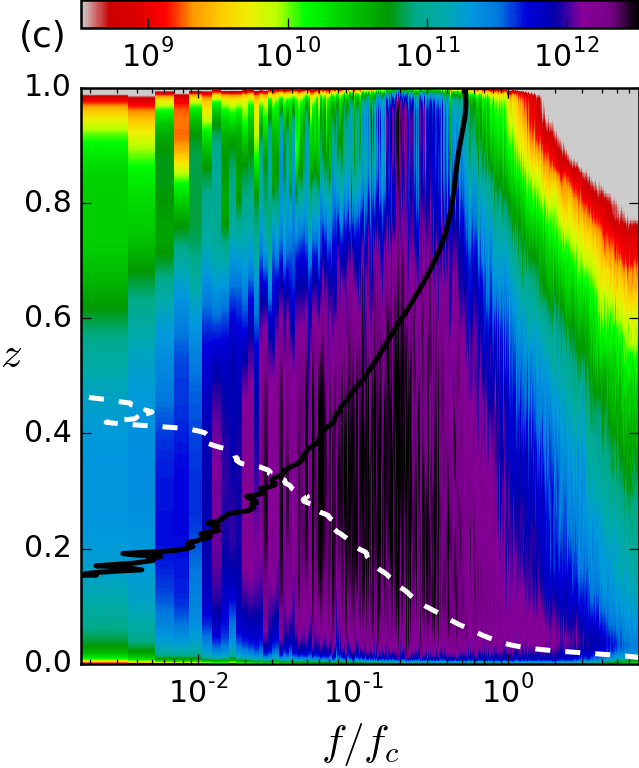}
\vspace{-.0in}
\caption{Spectrogram of the kinetic energy density spectrum $0.5 f \langle \hat{u}^2 + \hat{w}^2 \rangle_x$ as a function of frequency and $z$  for each of the three simulations shown in figure \ref{schematic} ($\Ra=8\times 10^7$; $\s=2^8,2^0,2^{-8}$ from left to right). The frequencies are normalized by the reference turnover frequency $f_c=\sqrt{\Ra\Pr}/2\pi$. The black solid (white dashed) line represents the mean value of the real (imaginary) part of the normalized buoyancy frequency, i.e. $N/f_c$ with $N=\overline{\langle \sqrt{-\Ra\Pr\p_z\rho} \rangle}_x/2\pi$ based on our notations. The arced dashed lines in (a) highlight global internal-wave modes.}\label{spectra}
\end{figure}

The regimes of entrainment- and plume-dominated heat transfer exhibit different types of flows and dynamical signatures. Figure \ref{spectra} shows the temporal kinetic energy density spectrum averaged in $x$ ($\langle\rangle_x$ operator), i.e. $2\pi f\langle K \rangle_x  = \pi f \langle \hat{u}^2(f,x,z) + \hat{w}^2(f,x,z) \rangle_x$, for the three simulations shown in figure \ref{schematic} ($\hat{\cdot}$ denotes temporal Fourier transform, $f$ the frequency).

The kinetic energy density spectrum for the high-stiffness case (figure \ref{spectra}a) shows a convective layer with a mid-plane symmetry, similar to classical RBC. As expected, the energy peaks at frequency $f/f_c\sim 1$, where $f_c=\sqrt{\Ra\Pr}/(2\pi)$ is the reference turnover frequency. The kinetic energy density decreases rapidly above the mean neutral buoyancy level ($z\sim 0.33$, see figure \ref{schematic}a) in the stably stratified region. Diffusive and viscous damping is most rapid for low-frequency internal waves (such as those at the convection frequency), so only relatively high-frequency waves carry energy all the way to the top of the domain. The kinetic energy density is small above the buoyancy frequency, shown as the black solid line. Interestingly, the arced dashed lines in figure \ref{spectra}a, which highlight low-energy $(z,f/f_c)$ paths, indicate the presence of nodes of standing high-frequency internal waves.

Figure \ref{spectra}b shows the kinetic energy density spectrum for the moderate-stiffness case. As for figure \ref{spectra}a, the energy density peaks at $f/f_c\sim 1$ in the convective region, but also at $f/f_c \sim 0.04$. A second peak at lower frequency can also be seen for the high-stiffness case (figure \ref{spectra}a), though less intense, that is due to a large-scale circulation, which is stable on very long timescales. For the moderate-stiffness case, the second peak may be due to the large-scale circulation, but is also possibly related to the coupling of the convective and wave dynamics; it may be due to the slow motion downward of low-density-anomaly stable fluid entrained by the penetrating plumes. The buoyancy frequency (black solid line) in the moderate-stiffness case is relatively close to the reference turnover frequency $f_c$, such that the kinetic energy density of the internal-wave field that peaks at $f/f_c\sim \co(1)$ quickly decreases with frequency. Compared to the high-stiffness case, internal waves are yet more strongly generated as a result of the penetrating plumes, complementing the generation by Reynolds stress due to weaker plumes sweeping beneath the neutral buoyancy interface (which is the main generation mechanism for stronger stratification, see \cite{Lecoanet2015}). We note that in the case of water cooled from below ($T_b^*=0\degree$C) and heated from above ($T_t^*=25\degree$C), the kinetic energy density spectrum obtained shows similarities with both figure \ref{spectra}a and \ref{spectra}b. This is because the stiffness parameter approximating the quadratic EoS for water (with $T_i^*=4\degree$C) is then $\s \approx (T_t^*-T_i^*)/(T_i^*-T_b^*) = 5.25$.

The kinetic energy density spectrum for the low-stiffness case ($\s=2^{-8}$) is shown in figure \ref{spectra}c. Fluid motions in the upper stable layer ($z\geq 0.8$) have a lot of energy at frequencies larger than $N$, showing that even though the stratification is stable, the motions are not wave-like. This is in part due to the fact that the stratification is weak, such that large convective motions from the entrained fluid are not strongly affected by the restoring buoyancy forces. The maximum of kinetic energy density spectrum in the bulk is approximately two orders of magnitude smaller for $\s=2^{-8}$ compared to the high-stiffness case. This is because buoyancy-driven fast plumes are confined near the bottom boundary (the energy peak is close to 1 near $z=0$), while slower larger-scale structures of lower frequencies dominate the fluid bulk (as already suggested from the large energy peak at $\sim 0.1$ in figure \ref{spectra}b); this can also be seen from the decrease in the frequency of the kinetic energy density maximum from $\sim 1$ close to the bottom boundary layer to $\sim 0.1$ at $z\sim 0.2$.

%%%%%%%%%
\subsection{Probability density functions}\label{sec44}
%%%%%%%%%

Probability density functions (PDFs) of flow variables describe important aspects of the flow characteristics. In experiments, PDFs of the flow variables are accessible using thermometers and PIV techniques, and geophysicists can infer flow statistics for e.g. Earth's deep interior from measurements of the magnetic field. Here we show that different convective regimes exhibit different flow statistics.

\begin{figure}
\centering
%\hspace{-1.2in}
\includegraphics[width=0.6\textwidth]{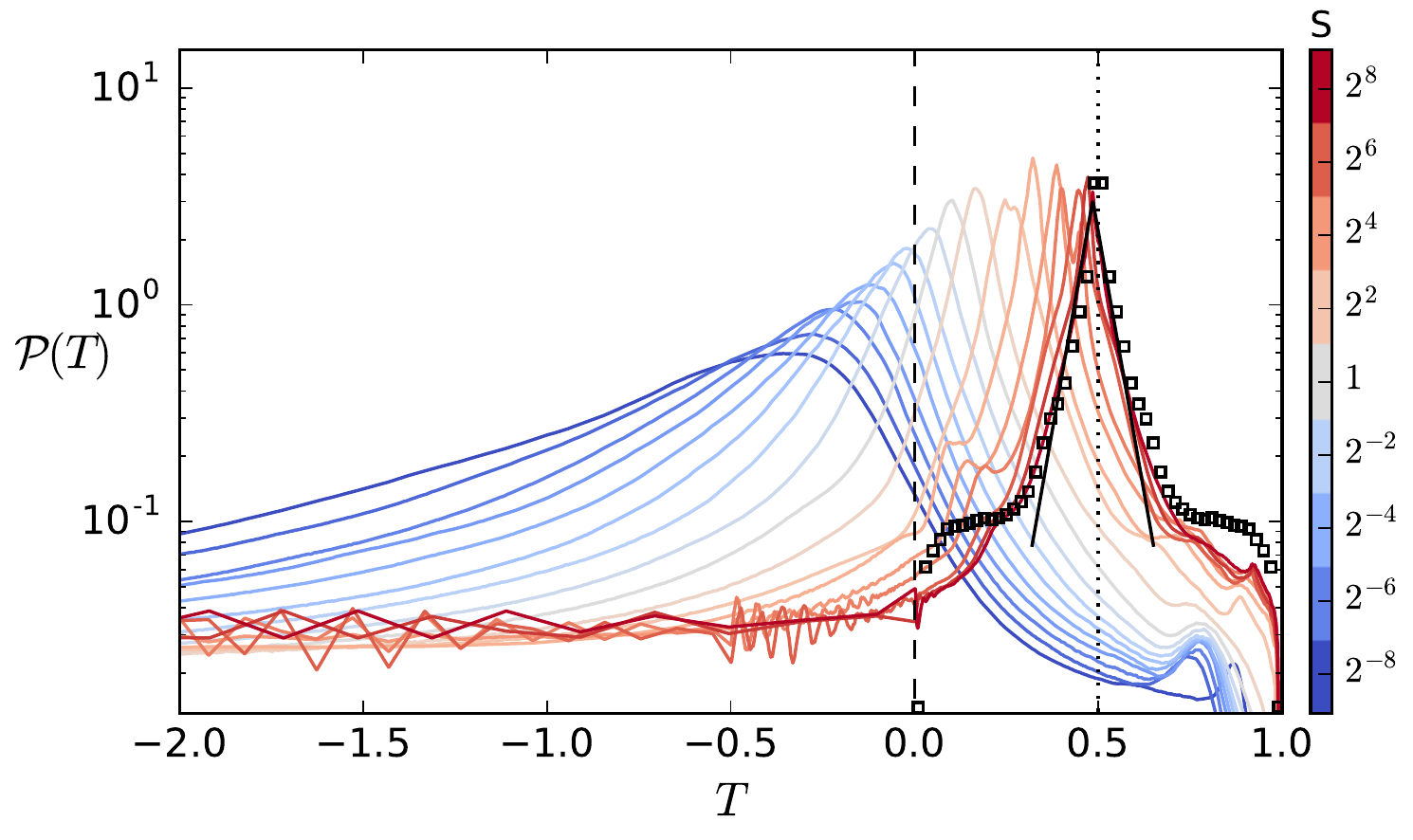}
\vspace{-.1in}
\caption{Probability density function  of temperature $\mathcal{P}(T)$ for different stiffness parameters $\s$ and $\Ra=8\times 10^7$. As $\s$ is increased, the peak of $\mathcal{P}(T)$ tends to 0.5 (dotted line), the bulk temperature expected for classical RBC. As $\s$ is decreased, the peak of $\mathcal{P}(T)$ drops below the inversion temperature (dashed line). The profile of $\mathcal{P}(T)$ is symmetric about the peak for hight stiffness (black solid line shows exponential fit), but asymmetric for low stiffness. Small local maxima of $\mathcal{P}(T)$ in the range $T\in[0.5,~1]$ are expected to be representative of the plumes' temperature. As in figure \ref{mean}, the unfilled-square symbols show results from a convection-only simulation with a vertical height comparable to the convection depth of the highest-stiffness case. Note that $\mathcal{P}(T)$ for the convection-only results is normalized such that the integral value is equal to the integral value of $\mathcal{P}(T)$ for $\s=2^8$ over the range $T\in[0,1]$.}\label{pdft}
\end{figure}

We show in figure \ref{pdft} the PDFs of the temperature for all our simulations with $\Ra=8\times 10^7$. All PDFs are obtained based on temperatures interpolated on a uniform grid within the entire domain, and have been normalized such that the integral value $\int \mathcal{P}(T)dT=1$. As the stiffness increases and becomes large, the peak of the PDF tends toward $T\approx 0.5$ and becomes more and more symmetric about its peak value, as is the case for convection-only simulations (see PDF shown by the unfilled-square symbols in figure \ref{pdft}). Moreover, the PDFs for the high-stiffness cases appear mostly exponential, in agreement with earlier studies \cite[][]{Siggia1994a,Kerr1996}. As the stiffness $\s$ decreases, however, the peak shifts toward negative temperatures and the PDF becomes skewed toward the lower temperature values. The decrease for low stiffness of the peak temperature is in agreement with the low average temperature reported in figure \ref{mean}, and the asymmetry tells us that the large-scale convecting structures have densities only marginally smaller than 0 (the maximum density, see figure \ref{mean}b). This asymmetry is a feature of the entrainment regime, and is not due to sampling a subset of the total convective domain \cite[see e.g.][]{Emran2008}. Asymmetric temperature distributions have already been observed in laboratory experiments of convection in water \cite[see][]{Lebars2015}, in which case the asymmetry is most likely related to entrainment and mixing of the stable fluid in the convection zone. Note that the oscillations of the temperature distributions for high-stiffness cases are most likely due to statistics that are not yet fully converged in the stable layer.

\begin{figure}
\centering
%\hspace{-0.2in}
\includegraphics[width=0.9\textwidth]{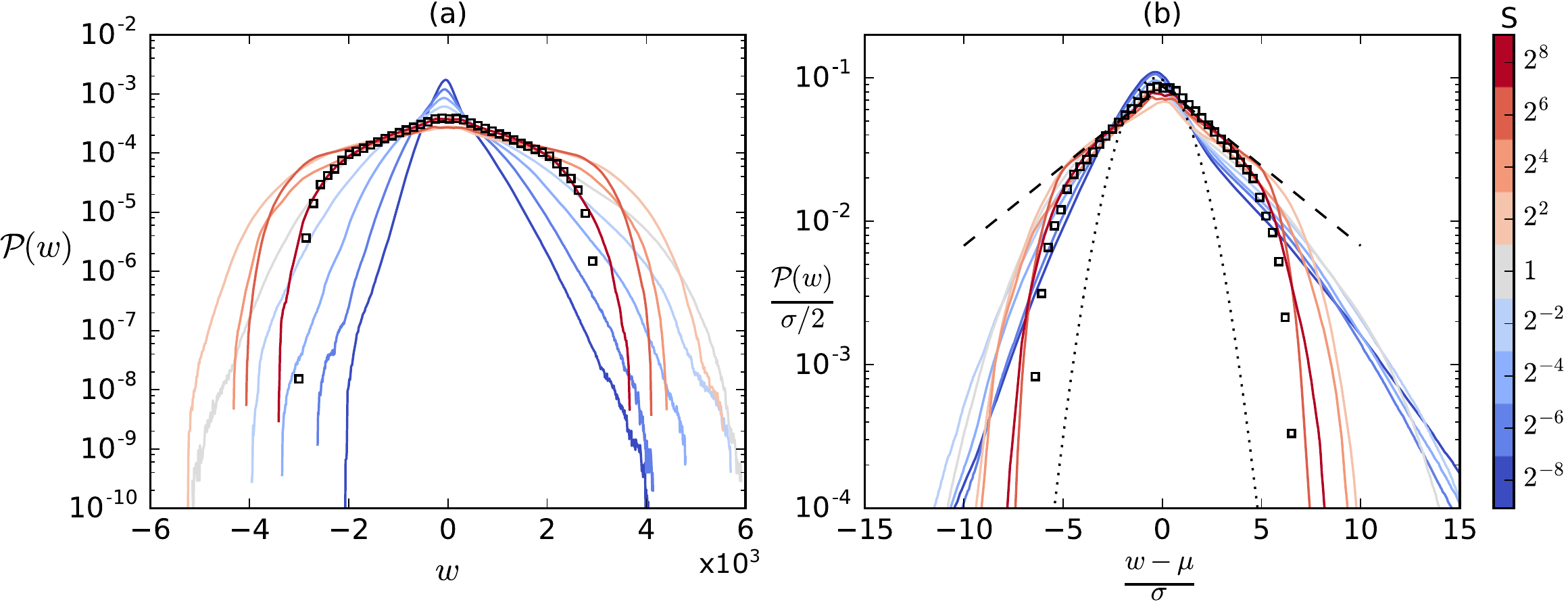}
\vspace{-.1in}
\caption{(a) Probability density function $\mathcal{P}$ of velocity $w$ from the convective region for different stiffness parameters $\s$ and for $\Ra=8\times 10^7$. Note that the convective region encompasses both the entrainment zone and the plume zone (cf. figure \ref{layers}). (b) Same as (a) but normalized by the standard deviation $\sigma$ and mean $\mu$ of the distributions. For high stiffness (red solid line), the distribution looks exponential for slow speeds (dashed line). For low stiffness (blue) the distribution looks Gaussian (dotted line) for slow speeds with asymmetric exponential tails. As in figure \ref{mean}, the unfilled-square symbols show results from a convection-only simulation with a vertical height comparable to the convection depth of the highest-stiffness case.}\label{pdfw}
\end{figure}

The PDFs of the vertical velocity $w$ obtained for different stiffnesses are shown in figure \ref{pdfw}a (for clarity, we do not include results from all simulations). The PDFs are for vertical velocities interpolated on a uniform grid within the convection zone only, and normalized such that $\int\mathcal{P}(w)dw=1$. As in classical RBC (cf. unfilled-square symbols), the PDF of $w$ for high stiffness is symmetric with respect to positive/negative values. The PDF for low stiffness is, on the other hand, skewed towards the positive vertical velocities $w>0$, and a similar albeit weaker asymmetry is also observed for $\s=2^0$. The asymmetry for $\s=2^{-8}$ is toward high positive vertical velocities because plumes, which are much faster than the entrained fluid, are only ascending. That the velocity distributions for the more flexible cases are more peaked around small velocities than for the stiffer cases suggests that patches of the stable fluid that are entrained within the convection zone have small velocities. In an attempt to further characterize the dynamics of the plume- and entrainment-dominated convection, we show in figure \ref{pdfw}b the PDFs of vertical velocity $w$ after normalization with respect to their mean and standard deviation. For the high-stiffness cases, we see that the slow speeds (in absolute magnitude) follow an exponential distribution while the distribution at higher speeds decays like a Gaussian (cf. dashed line). For the low-stiffness cases the result is the opposite: slow speeds seem to follow a Gaussian distribution (cf. dotted line), while high speeds follow an asymmetric exponential distribution. The statistics for weak stratification suggest a mixed dynamics. Large-scale structures that are relatively slow (but with a Reynolds number $\sim 100$ still relatively large based on a length scale of $\sim 1$ and typical rms velocity of $\sim 100$) have a velocity distribution similar to isotropic random fluctuations, possibly due to the fact that the  motions of the entrained fluid are not driven by buoyancy and thus are isotropic. Fast fluid motions (i.e. plumes rising in the dense fluid bulk), on the other hand, are intermittent and strongly skewed towards $w>0$. For the stiffest cases, fluid motions in the convection zone are expected to be turbulent, which is in agreement with the Gaussian distribution for the highest speeds. That the PDFs for the high-stiffness cases (and also the convection-only simulations) are not purely Gaussian for slow speeds may be explained from the fact that we have compiled statistics for $w$ not only in the bulk but also partly on the edge of the boundary layers, which are not as turbulent as the bulk.

\begin{figure}
\centering
%\hspace{-1.2in}
\includegraphics[width=0.32\textwidth]{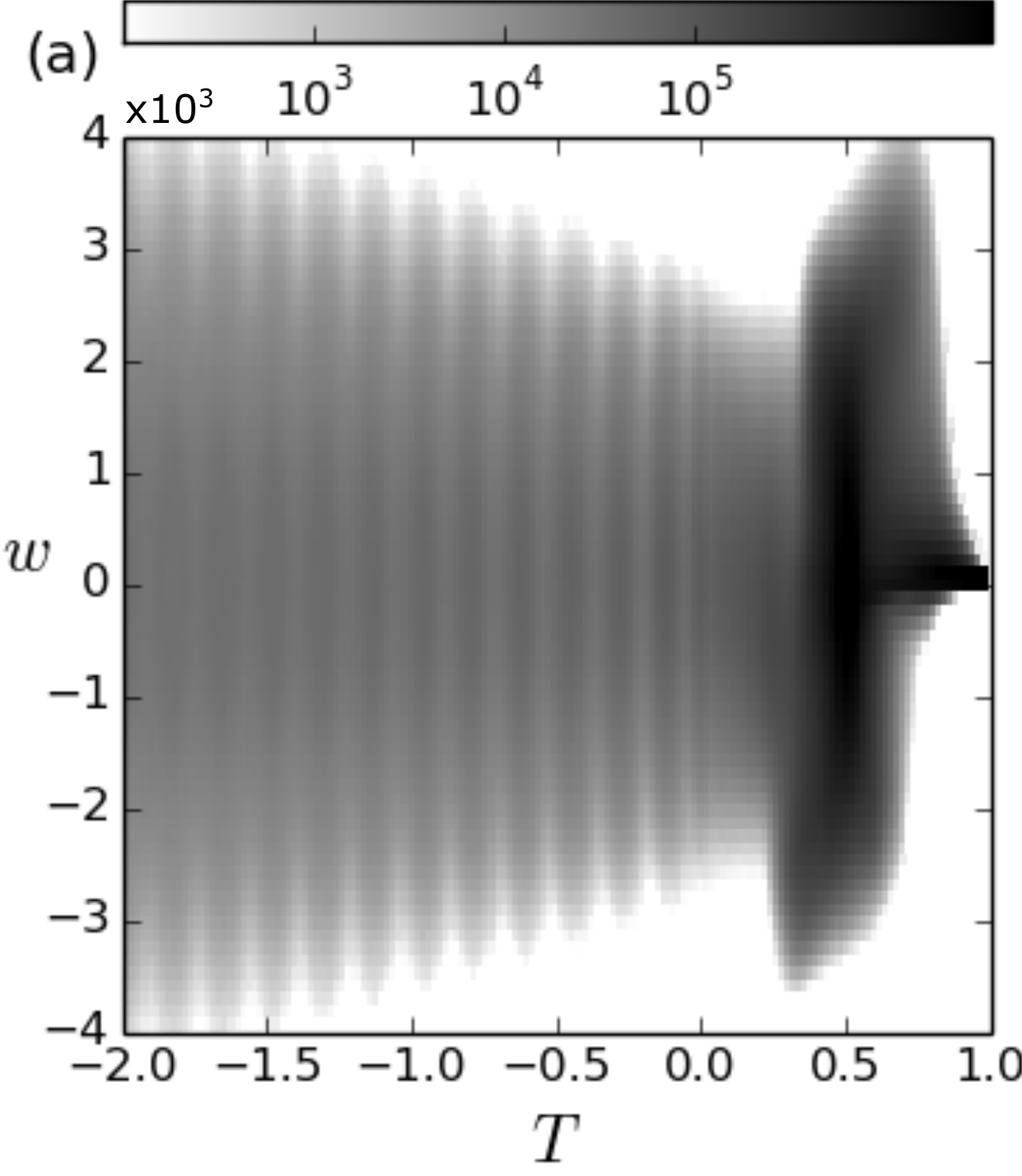}
\includegraphics[width=0.32\textwidth]{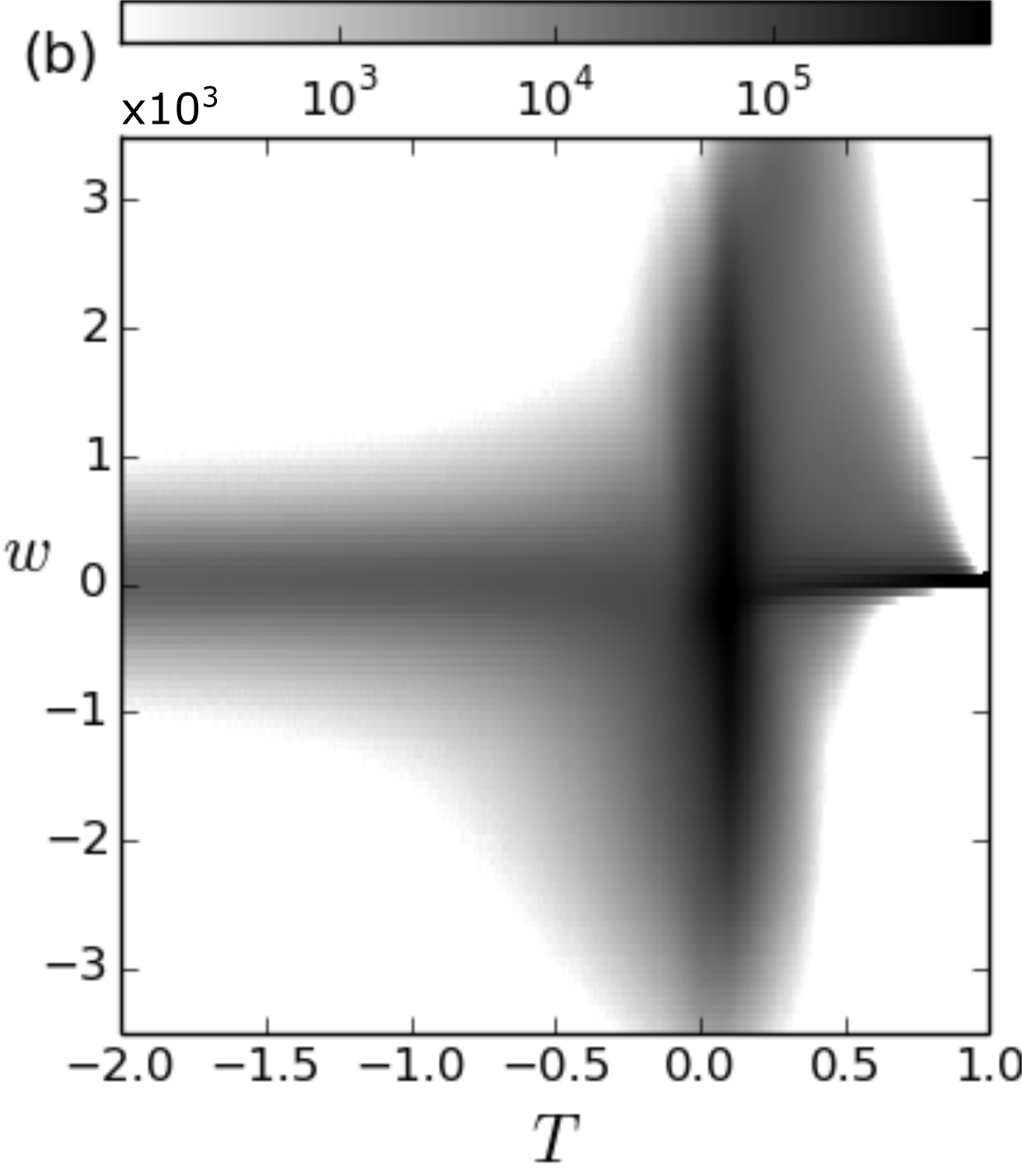}
\includegraphics[width=0.32\textwidth]{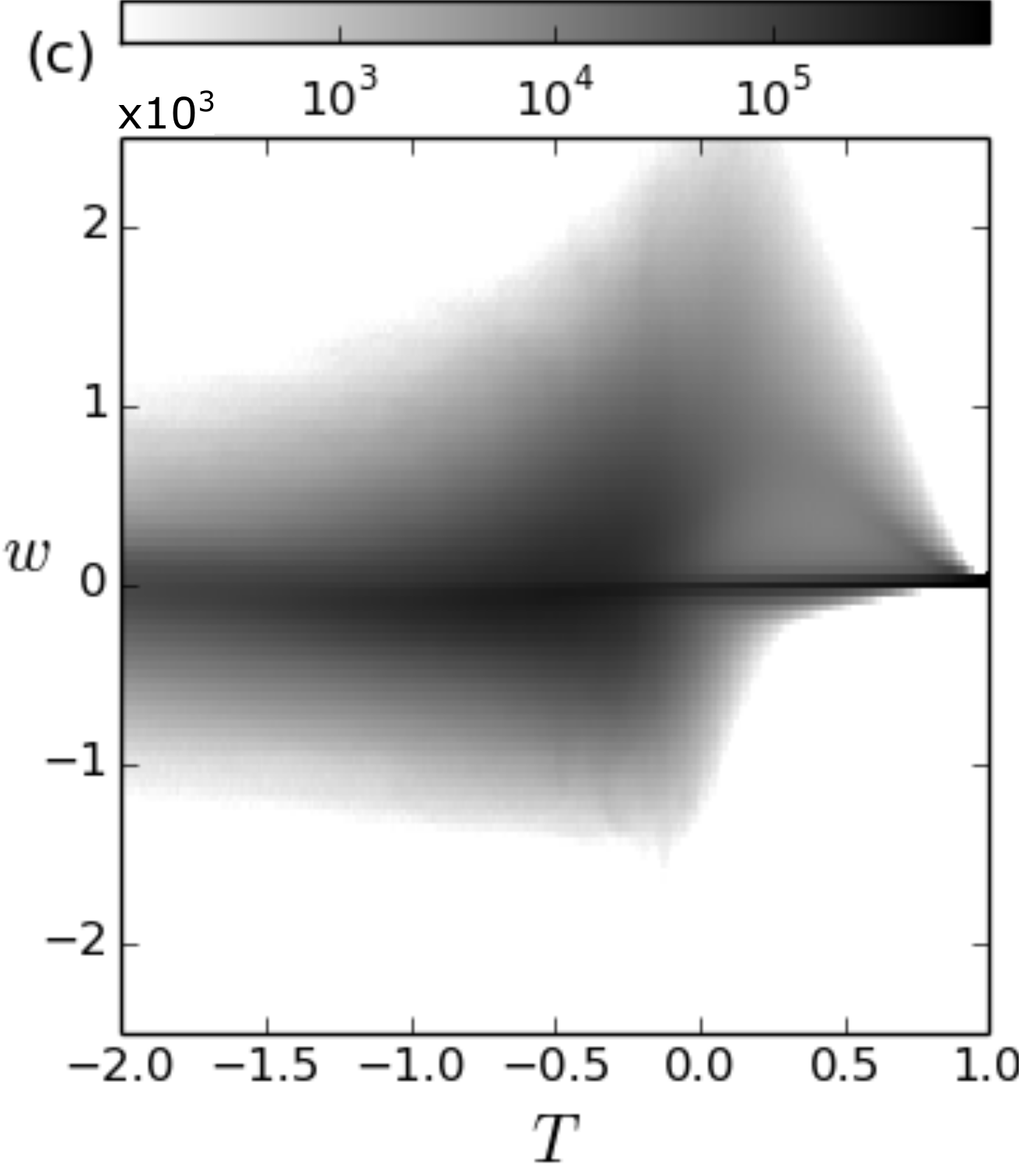}
\vspace{-.0in}
\caption{Joint probability density function of temperature and vertical velocity for (a) $\s=2^8$, (b) $\s=2^0$, (c) $\s=2^{-8}$ and $\Ra=8\times 10^7$. Note that the oscillations in (c) are most likely due to statistics that are not yet fully converged in the stable layer for the highest-stiffness case.}\label{joint}
\end{figure}

Figures \ref{joint}a-c show the joint PDFs of $w$ and $T$ for $\s=2^8,2^0,2^{-8}$. For large stiffness (figure \ref{joint}a), the joint PDF shows a symmetry for $T\in[0,1]$ around zero velocity and $T=0.5$, because the heat transfer is due to both rising and sinking plumes. In figure \ref{joint}b, the peak convective heat transfer occurs for values of $T$ closer to 0, and the larger values of the joint PDF for $T>0$ and $w>0$ highlight the increased importance of hot rising plumes compared to cool descending ones. In the weak-stratification regime (figure \ref{joint}c), low-velocity sinking fluid parcels are primarily responsible for the convective heat flux, in addition to the slightly-hot fast rising plumes (seen in figure \ref{schematic}c). The asymmetry of positive/negative vertical velocities is evident and again demonstrates that buoyancy-driven plumes only go upward.

%%%%%%%%
\subsection{Heat transfer scaling}\label{sec45}
%%%%%%%%

We now discuss the Nusselt-Rayleigh scaling obtained from our simulations, which is an essential element of the study of convective heat transfers. The Nusselt number $\Nu$ is traditionally defined as the total heat transfer $Q$, normalized by the purely conductive heat flux, and thus $\Nu$ measures the heat transfer enhancement due to convective motions. In our case, the purely conductive heat flux across the domain boundaries is $-(T_t-1)$, such that 
\ba{}\label{Nu1}
\Nu=\f{-Q}{T_t-1}
\ea
In classical RBC, it is generally assumed that one can write $\Nu(\Ra,\Pr)=C \Ra^{\beta}\Pr^{\gamma}$, with $\Ra$ and $\Pr$ the two (separable) problem parameters. Here, we have two additional parameters ($\s$ and $T_t$), but as we restrict $\Pr=1$ and $T_t=-20$, we only consider $\Nu\equiv\Nu(\Ra,\s)$.

\begin{figure}
\centering
%\hspace{1.0in}
\includegraphics[width=0.55\textwidth]{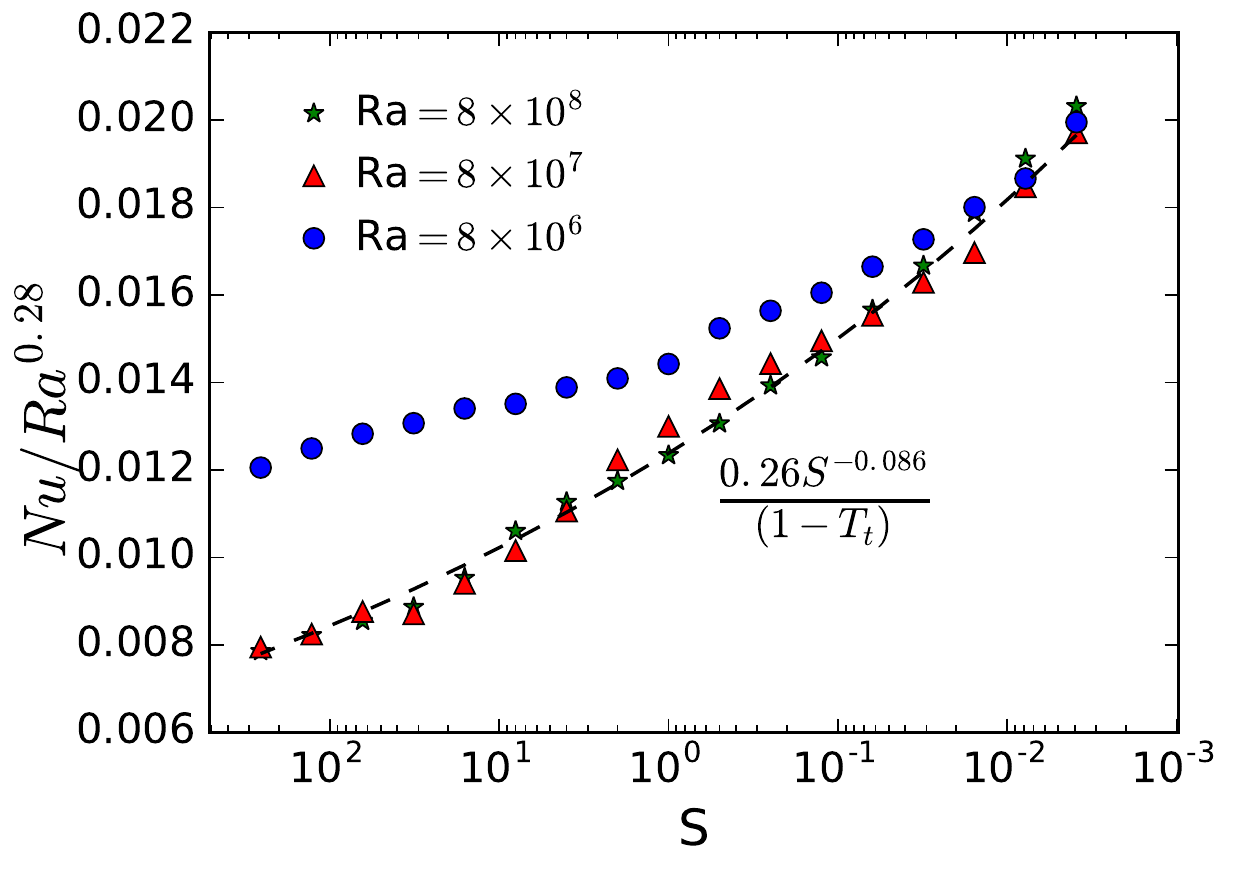}
\vspace{-.1in}
\caption{Variations of $\Nu/\Ra^{0.28}$ with $\s$ where $\Nu=-Q/(T_t-1)$ for three reference 
Rayleigh numbers. The dashed line shows the best fit power law regression of $\Nu$ in terms of $\s$ for $\Ra=8\times 10^7,8\times 10^8$.}\label{nura}
\end{figure}

Figure \ref{nura} shows $\Nu/\Ra^{0.28}$ as a function of $\s$ for all three reference Rayleigh numbers. The exponent $0.28$ for $\Ra$ is obtained from a best fit power law regression of $\Nu$ for our simulations with $\Ra=8\times 10^7,8\times 10^8$ for each $\s$, and is within the range obtained in classical RBC \cite[][]{Ahlers2009}. We do not expect the RBC scaling to work equally well for all $\s$, and the r$^2$-value of this first fit is therefore small (i.e. $\sim 0.52$) due to the relatively large dispersion of the exponents in the range $[0.26,0.31]$ obtained for different $\s$. Nevertheless, there is then a relatively good collapse of the data for different $\s$ with exponent $-0.086$ (with final r$^2$-value of 0.99). That the exponent is negative is expected since as the stiffness decreases we showed that the heat transfer increases. Note, however, that this exponent has no theoretical grounding yet. For the smaller $\Ra=8\times 10^6$, $\Nu/\Ra^{0.28}$ is above the collapsed data, especially for the high-stiffness cases, because the convection pattern is steady. This is consistent with classical RBC studies which showed that the Nusselt number has a steeper scaling with $\Ra$ for smaller $\Ra$, close to the onset of convection.

The empirical law 
\ba{}\label{Nu3}
\Nu(\Ra,\s)\approx  \f{0.26 \Ra^{0.28}\s^{-0.086}}{(1-T_t)}
\ea
suggested by figure \ref{nura} can be useful in predicting the size of the convection zone in mixed convective--stably-stratified systems. However, the Nusselt numbers  obtained are small (we find $\Nu\in [1,7]$), compared to those demonstrated in classical RBC, where Nusselt numbers have been reported in the range $\Nu \in [10,30] $ for $\Pr=1$ and Rayleigh numbers in $8\times 10^6$--$8\times 10^8$ \cite[see e.g.][]{Grossmann2000}. In order to reconcile the expected enhancement of the heat transfer due to convective motions in our simulations in the high-stiffness limit with RBC results, we consider the rescaled Nusselt number (subscript $C$ standing for convection)
\ba{}\label{Nu2}
\Nu_{C}=\f{Q}{1/h},
\ea
where $h=(2L_{db}+L_{cp}+L_{ce})$ is a measure of the depth of the convection zone based on \eqref{a13}. $\Nu_C$ thus corresponds to the heat flux normalized by the diffusive heat flux through the depth of the convective layer only.

\begin{figure}
\centering
%\hspace{1.0in}
\includegraphics[width=0.6\textwidth]{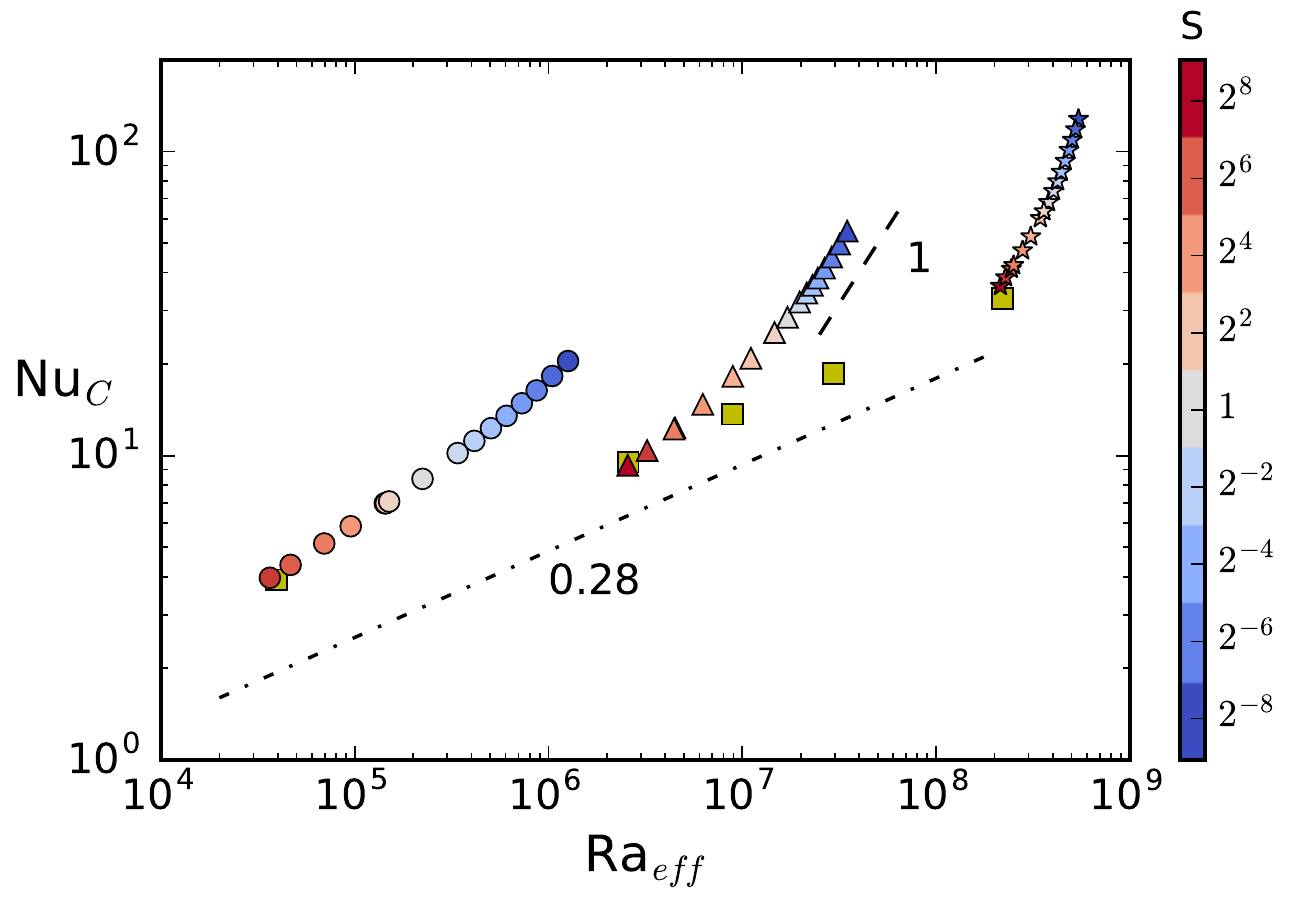}
\vspace{-.1in}
\caption{Rayleigh-Nusselt scaling based on $\Nu_C=Qh$ and effective Rayleigh number $\Ra_{eff}=\Ra h^3$ where $h$ is an estimate of the convective layer depth (see equation \eqref{Nu2} and  text for details). The circles, triangles and stars correspond to reference Rayleigh number $\Ra=8\times 10^6,8\times 10^7,8\times 10^8$, respectively. The dash-dot line has slope 0.28, while the dashed line has slope 1. Results of five convection-only simulations are shown by filled squares; note that the lowest $\Ra_{eff}$ convection-only simulation is stationary in time; also, the scaling exponent for the convection-only results is 0.28 (discarding the lowest $\Ra_{eff}$ case)% with r$^2$-value 0.998
.}\label{nure}
\end{figure}

In figure \ref{nure} we plot $\Nu_C$ as a function of the effective Rayleigh number $\Ra_{eff}=\Ra h^3$ (also characterizing this convective layer only). We find that $\Nu_C \approx C\Ra_{eff}^{0.28}$ for each fixed $\s$ where $C$ is a constant that depends on $\s$. For the high-stiffness cases ($\s=2^8$) we find that $\Nu_C$ is in the range 3--30 for $\Ra_{eff} \in [2\times 10^4,2\times 10^8]$, in relatively good agreement with classical RBC results (shown by the filled squares). For fixed reference $\Ra$ number, $d\log\Nu_C/d\log(\Ra_{eff})$ appears to approach $0.28$ as $\s \rightarrow \infty$, similar to classical RBC, which suggests that  the coupling between the two layers diminishes. For small stiffness, however,  $d\log\Nu_C/d\log(\Ra_{eff})$ increases up to 1 for $\Ra=8\times 10^7$ and $\s=2^{-8}$, indicating that the weakly stratified fluid significantly enhances the convective heat transfer.

%%%%%%%%
\subsection{Extension of Moore's model to include the effect of $\s$}\label{sec46}
%%%%%%%%

%We demonstrate that the heat transfer increases with decreasing stiffness, and that it can be relatively well estimated from the scaling law \eqref{Nu3} . 
With the help of \eqref{Nu3} we can now extend Moore's approximate model \cite{Moore1973} for the prediction of the convective layer depth. This is done by equating the diffusive heat flux in the stable region with the convective heat flux, i.e.
\ba{}\label{hpred0}
\f{-\lb T_t-T_{bulk}(\s) \rb}{1-h} \approx -\Nu(\s)(T_t-1),
\ea
with consideration of the effect of $\s$ on $\Nu(\s)$ and $T_{bulk}(\s)$ (approximately the temperature at the top of the convective layer) through equations \eqref{Nu3} and \eqref{tfit}. We obtain
\ba{}\label{hpred}
h \approx 1-\f{\lb T_{bulk}(\s)-T_t \rb}{0.26\Ra^{0.28}\s^{-0.086}},
\ea
and the prediction \eqref{hpred} (with $T_{bulk}$ substituted by \eqref{tfit}) is shown in figure \ref{layers} by the filled diamonds. It should be noted that in the high-stiffness limit $\Nu$, $T_{bulk}$ and $h$ seem to become independent of $\s$, so \eqref{hpred} cannot be valid as $\s\rightarrow \infty$. That some variations of $\Nu$ are observed for $\s\in[2^6,2^8]$ shows that we are not yet in the high-stiffness regime even with $\s=2^8$, and suggests that RBC is quantitatively recovered only when $\s \gg \co(100)$. The classical RBC scaling of heat transfer is nevertheless recovered for the highest stiffness considered, when choosing an appropriate effective Rayleigh number (see figure \ref{nure}).

%%%%%%%%%%%%%%
\section{Conclusions}\label{sec5}
%%%%%%%%%%%%%%

We have presented two-dimensional direct numerical simulation results of thermal convection with buoyancy reversal, and demonstrated that the convective and internal-wave dynamics can be strongly coupled depending on the stable stratification strength. Our results are expected to hold qualitatively for arbitrary profiles of the thermal expansion coefficient with temperature, despite only considering a simple piecewise linear EoS. Similarities with convection in water, for which the EoS is quadratic, have been demonstrated by comparing our results with previous numerical and laboratory experiments \cite{Lecoanet2015,Lebars2015}.

\begin{figure}
%\centering
\hspace{.3in}
\includegraphics[width=0.80\textwidth]{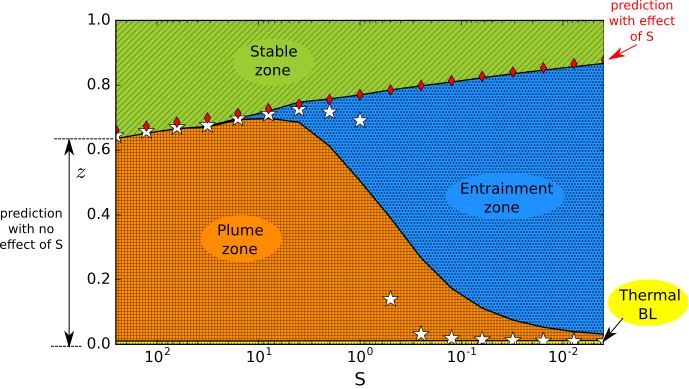}
\vspace{-.0in}
\caption{Same as figure \ref{layers} but for the higher Rayleigh number $\Ra=8\times 10^8$ case.}\label{supp8eRa}
\end{figure}

The different dynamical regimes obtained for different $\s$ are best summarized by figure \ref{layers} (see also figure \ref{supp8eRa} for higher Rayleigh simulations), and the two most important findings are that (i) for weak stratification ($\s\leq\co(10^{-1})$), entrained fluid motions that are maintained by rising plumes close to the bottom boundary dominate the convection zone, and (ii) classical RB convection is recovered only for $\s\gg 100$ since otherwise the influence of the stable layer on the mean temperature profiles and flow statistics is significant. For the smallest stiffnesses considered, the entire stable layer slumps down and becomes entrained by the buoyant plumes, such that the convective dynamics is qualitatively very different from what is expected from classical RBC or even penetrative convection. As $\s$ increases, however, entrained fluid motions vanish and the convective dynamics becomes again dominated by buoyant plumes; as expected, the overshooting length scale and mixing of the stable fluid decreases with increasing $\s$. For large stratification (i.e. $\s\geq\co(100)$), the convection is relatively similar to RBC, the convective layer depth can be estimated assuming no feedback from the stably-stratified fluid (see equation \eqref{a55}), and flow statistics are similar to convection-only statistics.

In the high-stiffness limit, the convection is not affected much by the waves it generates, such that classical RBC statistics might be used as forcing in simulations of internal waves within the stable zone. Whether the theoretical approach in \cite{Lecoanet2013} or numerical method based on Reynolds stress \cite[][]{Lecoanet2015} is valid is, however, beyond the scope of this manuscript. In fact, as $\s$ decreases to $\s\leq\co(10)$, Reynolds stress may not be the only mechanism exciting internal waves, so it may be necessary to consider other mechanisms such as the mechanical oscillator effect \cite[see e.g.][for a discussion of the mechanical oscillator effect in relation with the generation of atmospheric internal waves by storms]{Ansong2010}.

We have limited our study to fixing the top temperature to $T_t=-20$, which results in approximately equal-size convective and stable layers for many parameter values.  The effects of changing $T_t$ are qualitatively predictable and are not expected to change the importance of the convective/wave coupling. Imagine a domain with larger $T_t$ (e.g., $T_t=-10$). If the convection height is the same as for the case $T_t=-20$, the temperature gradient in the stable layer is larger (i.e. less negative), such that the heat flux is smaller in the stable than in the convective layer. To obtain a thermal equilibrium the convection height $h$ must thus be different. In particular, $h$ should decrease as $T_t<0$ decreases (in agreement with equation \eqref{a55}), because the diffusive heat flux of the upper layer increases with $h$ (i.e. $Q_d\approx -T_t/(1-h)$) while the convective heat flux of the lower layer stays the same (at least in the high-stiffness case). Preliminary results for $\s=1$ have confirmed this claim. For $T_t=-20$ we have $h\approx 0.60$ and $Q \approx 48$, whereas for $T_t=-40$ we find $h\approx 0.27$ and $Q \approx 55$, and for $T_t=-10$, we find $h\approx 0.89$ and $Q\approx 44$. Note that the increase of $Q$ with decreasing $T_t$ suggests that the effect of $T_t$ on $h$ is not quite as dramatic as indicated by \eqref{a55}, at least for $\s=1$.

The effect of the Prandlt number, which we did not explore (we set $\Pr=1$), is likely to be more complex than $T_t$ and will be reported elsewhere. Nonetheless, we might expect that decreasing $\Pr$ will result in stronger convective/wave coupling. Plumes overshooting from the convective region are expected to penetrate further into the stratified layer as a result of rapid thermal diffusion processes in the low P\' eclet limit. Significant overshooting could potentially result in turbulent motions within the stably-stratified layer, a topic of interest to astrophysical systems that would be worth investigating \cite[][]{Brummell1993,Garaud2015}.

We have based our analysis on the stiffness parameter $\s$, which is equal to the ratio of the thermal expansion coefficients, or $N^2/(dT/dz)$, in the stable and convective regions. In real systems, we can only measure the end states of nonlinear processes, which means that $N^2$ and $dT/dz$ are expected to be both approximately zero in the well-mixed (convective) zone and hence difficult to measure. Extracting $\s$ from real systems might thus require knowing the thermal expansion coefficient within the convection based on the fluid composition directly, while measuring both $N^2$ and $dT/dz$ in the stably-stratified zone. A different stiffness parameter, $\s_H$, has been used in previous astrophysical studies of compressible convection \cite{Hurlburt1994}. Anelastic simulations have a background temperature profile $\bar{T}(z)$, and $\s_H$ is defined as the ratio of $N^2/(dT/dz)$ in the two regions, multiplied by the ratio of $\bar{T}(z)$ in different parts of the domain. For domains smaller than a typical pressure scale height (the Boussinesq regime), $\bar{T}(z)$ is almost constant, and thus $\s \approx \s_H$. Massive stars have stable layers with typical stiffness $\s_H\sim 10^5$, and therefore are good examples of coupled systems in which the stable layer has a large stiffness \cite{shiode2013}. Knowing whether the convection is similar to RBC or include some types of entrained motions would yet still require a self-consistent study, because the Rayleigh number in stars is so large that the required stiffness for the entrainment to be negligible should be very large too. In the case of Earth's liquid outer core, the stratification strength of the stable layer--if it exists--is still uncertain (possibly not too different from the Coriolis frequency, \cite[][]{Vidal2015}) and the Rayleigh number is relatively large \cite[][]{gubbins2001}. Therefore, self-consistent studies might be required in the context of planetary interiors in order to estimate the importance of the stable layer's feedback on the convection. It is possible that convective motions in the oceans and atmosphere \cite{marshall1999,parodi2009} may sometimes fall within the class of low-to-moderate stiffness convective regimes (i.e. including some entrainment), depending on the properties of the stably-stratified fluid \cite{grise2010,king2012}.

The plume- and entrainment-dominated regimes have been shown to have unique dynamical signatures that can be of interest to astrophysicists, meteorologists, oceanographers, and geophysicists. Understanding how the flow signatures relate to the mean state of a fluid is of significant importance since measurements of, e.g., the stable stratification in planets and stars are typically difficult, and the present study can provide information on the dynamics of a system with a buoyancy reversal. We have neglected a number of physical effects that can have a significant impact on astrophysical and geophysical fluids. In the case of stars and planetary interiors, the mean background convective state does not have a constant density profile, but a constant entropy profile, and compressibility and magnetic effects can be significant. Non-uniform background density profiles can lead to upward buoyancy plumes stronger than downward ones (and vice-versa), and strong variations of the buoyancy frequency even deep in the stable region, resulting in internal waves of varying amplitudes \cite{dosser2011}. Double diffusive effects may also play a major role in some cases. In the ocean, for instance, density is a function of both temperature and salt such that temperature gradients can be stabilizing while salt stratification is destabilizing, and vice-versa. Our generic model of convective--stably-stratified fluids provides a basis for future works aimed at considering such physical effects, but also spherical geometry and rotation, along with buoyancy reversal. The potential emergence of a mean flow in the stable layer similar to the Quasi-Biennial Oscillation of the Earth's equatorial stratosphere \cite{Baldwin2001} is also a topic of significant importance which will be explored in the near future.

\section*{Acknowledgement}

This research is funded by the European Research Council (ERC) under the European Union’s Horizon 2020 research and innovation program through grant agreement No 681835-FLUDYCO-ERC-2015-CoG. D.L. is supported by a PCTS fellowship, and a Lyman Spitzer Jr.~fellowship. This project has been carried out in the framework of the Labex MEC (ANR-10-LABX-0092) and of the A*MIDEX project (ANR-11-IDEX-0001-02), funded by the program ``Investissements d’Avenir'' supervised by the Agence Nationale de la Recherche. This work was granted access to the HPC resources of Aix-Marseille Universit\' e financed by the project Equip@Meso (ANR-10-EQPX-29-01) of the program  ``Investissements d'Avenir'' supervised by the Agence Nationale de la Recherche, and to the HPC ressources of IDRIS (clusters Turing and Ada) under allocation A0020407543. Additional computing resources supporting this work were provided by the NASA High-End Computing (HEC) Program through the NASA Advanced Supercomputing (NAS) Division at Ames Research Center. This project was supported in part by NASA under TCAN grant number NNX14AB53G, and by the National Science Foundation under Grant No. NSF PHY-1125915.

%%%%%%%%%%%%
\section*{Appendix A: Numerical method and simulation details}
%%%%%%%%%%%%

We use the open-source pseudo-spectral code Dedalus \cite{Burns2017} to solve equations \eqref{a3}. The spectral decomposition consists of Chebyshev polynomial functions in $\hat{z}$ direction and complex exponential functions (Fourier series) in the periodic $\hat{x}$ direction. We note $n_z$ and $n_x$ the number of modes in the $\hat{z}$ and $\hat{x}$ directions, respectively. The use of Chebyshev polynomials along the vertical axis results in increased resolution near the top and bottom domain boundaries, such that an accurate treatment of the thermal and viscous boundary layers is obtained with relatively few spectral modes. A 2-step implicit/explicit Runge-Kutta scheme is used for time integration \cite[][]{Ascher1997}.  We list all simulation runs whose results are presented in this paper in table \ref{table} with relevant numerical and physical parameters.

In order to reach thermal equilibrium relatively rapidly, we initiate all numerical simulations  at $t=0$ with low-amplitude noise added to a background zero-velocity field ($\u=\b{0}$) and a background temperature field of the form (except those in section \ref{sectransients}, which are started from the conductive state): 
\bsa{}\label{back}
T& = T_b + (T_{bulk}-T_b) \f{z}{z_{conv}} , ~~z \leq z_{conv}, \\
T& = T_t + (T_{bulk}-T_t) \f{1-z}{1-z_{conv}}, ~~ z> z_{conv}.
\esa
In equation \eqref{back} $z_{conv}$ is an estimate of the convective layer depth and $T_{bulk}$ the bulk temperature both obtained from preliminary simulations at low resolution, which allow us to initialize the problem close to thermal equilibrium. An accurate estimate for $z_{conv}$ and $T_{bulk}$ results in an accurate estimate for the heat flux $-(T_t-T_{bulk})/(1-z_{conv})$, and hence the temperature field in the stably-stratified layer, which is the fluid region that is longest to reach equilibrium.  While the convection reaches a dynamical equilibrium relatively quickly (i.e.~in $\sim \co(100)$ turnover times, which corresponds to $\co(0.01)$ thermal time for $\Ra=10^8$), the stably-stratified fluid layer above needs $\sim \co(1)$ thermal time to reach equilibrium. Setting up the initial temperature field close to thermal equilibrium in the stable region thus reduces the simulation time significantly by allowing us to obtain converged flow statistics rapidly, i.e. without having to wait several thermal times for the entire system to be at equilibrium. We note that the initial state of the system is of much less importance for a convective system without buoyancy reversal since in this case the transient initial phase consists of an exponential growth of the most unstable modes until rapid saturation.

We would like to note that the cadence at which data must be output is controlled by the shortest of the convective time scale $\tau_c=2\pi/\sqrt{\Ra\Pr}$ and the internal-wave time scale $\tau_w=\tau_c/\sqrt{\s}$. In order to construct flow statistics that capture the fastest dynamics here we typically use $\min(\tau_c,\tau_w)/10$ as the time step between two data outputs. It should be noted that when the output cadence is high ($\s$ large), statistics are at equilibrium with relatively short time averages. When the output cadence is long, however, longer time averages are necessary to ensure statistical steady-state. This implies that about the same number of data outputs are required to obtain converged statistics for all parameters $\s$.

\begin{table}
\centering
	\begin{tabular}{|c|c|c|c|c|}
	\hline 
$\Ra$		& $\s$ 		& $n_x \times n_z$ 	& CFL & dt	\\ \hline	 
$8\times 10^6$ & $2^{-8}$ & $256 \times 256$ 	& 0.7 & $\co(3\times 10^{-6})$ \\
$8\times 10^6$ & $2^{-7}$ & $256 \times 256$ 	& 0.7 & $\co(3\times 10^{-6})$ \\
$8\times 10^6$ & $2^{-6}$ & $256 \times 256$ 	& 0.7 & $\co(3\times 10^{-6})$ \\
$8\times 10^6$ & $2^{-5}$ & $256 \times 256$ 	& 0.7 & $\co(3\times 10^{-6})$ \\
$8\times 10^6$ & $2^{-4}$ & $256 \times 256$ 	& 0.7 & $\co(3\times 10^{-6})$ \\
$8\times 10^6$ & $2^{-3}$ & $256 \times 256$ 	& 0.7 & $\co(3\times 10^{-6})$ \\
$8\times 10^6$ & $2^{-2}$ & $256 \times 256$ 	& 0.7 & $\co(3\times 10^{-6})$ \\
$8\times 10^6$ & $2^{-1}$ & $256 \times 256$ 	& 0.7 & $\co(3\times 10^{-6})$ \\
$8\times 10^6$ & $2^{0}$ & $256 \times 256$ 	& 0.5 & $\co(10^{-6})$ \\
$8\times 10^6$ & $2^{1}$ & $256 \times 256$ 	& 0.5 & $\co(10^{-6})$ \\
$8\times 10^6$ & $2^{2}$ & $256 \times 256$ 	& 0.5 & $\co(10^{-6})$ \\
$8\times 10^6$ & $2^{3}$ & $256 \times 256$ 	& 0.5 & $\co(10^{-6})$ \\
$8\times 10^6$ & $2^{4}$ & $256 \times 256$ 	& 0.5 & $\co(10^{-6})$ \\
$8\times 10^6$ & $2^{5}$ & $256 \times 256$ 	& 0.5 & $\co(10^{-6})$ \\
$8\times 10^6$ & $2^{6}$ & $256 \times 256$ 	& 0.5 & $\co(10^{-6})$ \\
$8\times 10^6$ & $2^{7}$ & $256 \times 256$ 	& 0.5 & $\co(10^{-6})$ \\
$8\times 10^6$ & $2^{8}$ & $256 \times 256$ 	& 0.5 & $\co(10^{-6})$ \\ \hline
	\end{tabular}
	\caption{See caption below.}	
	\label{table}
\end{table}

\begin{table}
\ContinuedFloat
\centering
	\begin{tabular}{|c|c|c|c|c|}
	\hline 
$\Ra$		& $\s$ 		& $n_x \times n_z$ 	& CFL & dt	\\ \hline	 
$8\times 10^7$ & $2^{-8}$ & $512 \times 256$ 	& 0.7 & $\co(7\times 10^{-7})$ \\
$8\times 10^7$ & $2^{-7}$ & $512 \times 256$ 	& 0.7 & $\co(7\times 10^{-7})$ \\
$8\times 10^7$ & $2^{-6}$ & $512 \times 256$ 	& 0.7 & $\co(7\times 10^{-7})$ \\
$8\times 10^7$ & $2^{-5}$ & $512 \times 256$ 	& 0.7 & $\co(7\times 10^{-7})$ \\
$8\times 10^7$ & $2^{-4}$ & $512 \times 256$ 	& 0.7 & $\co(7\times 10^{-7})$ \\
$8\times 10^7$ & $2^{-3}$ & $512 \times 256$ 	& 0.5 & $\co(3\times 10^{-7})$ \\
$8\times 10^7$ & $2^{-2}$ & $512 \times 256$ 	& 0.5 & $\co(3\times 10^{-7})$ \\
$8\times 10^7$ & $2^{-1}$ & $512 \times 256$ 	& 0.5 & $\co(3\times 10^{-7})$ \\
$8\times 10^7$ & $2^{0}$ & $512 \times 256$ 	& 0.5 & $\co(3\times 10^{-7})$ \\
$8\times 10^7$ & $2^{1}$ & $512 \times 256$ 	& 0.5 & $\co(3\times 10^{-7})$ \\
$8\times 10^7$ & $2^{2}$ & $512 \times 256$ 	& 0.5 & $\co(3\times 10^{-7})$ \\
$8\times 10^7$ & $2^{3}$ & $512 \times 256$ 	& 0.5 & $\co(3\times 10^{-7})$ \\
$8\times 10^7$ & $2^{4}$ & $512 \times 256$ 	& 0.5 & $\co(3\times 10^{-7})$ \\
$8\times 10^7$ & $2^{5}$ & $512 \times 256$ 	& 0.35 & $\co(2\times 10^{-7})$ \\
$8\times 10^7$ & $2^{6}$ & $512 \times 256$ 	& 0.35 & $\co(2\times 10^{-7})$ \\
$8\times 10^7$ & $2^{7}$ & $512 \times 256$ 	& 0.35 & $\co(2\times 10^{-7})$ \\
$8\times 10^7$ & $2^{8}$ & $512 \times 256$ 	& 0.35 & $\co(2\times 10^{-7})$ \\ \hline
	\end{tabular}
	\caption{(Cont.) See caption below.}	
	\label{table}
\end{table}

\begin{table}
\ContinuedFloat
\centering
	\begin{tabular}{|c|c|c|c|c|}
	\hline 
$\Ra$		& $\s$ 		& $n_x \times n_z$ 	& CFL & dt	\\ \hline	 
$8\times 10^8$ & $2^{-8}$ & $512 \times 256$ 	& 0.5 & $\co(10^{-7})$ \\
$8\times 10^8$ & $2^{-7}$ & $512 \times 256$ 	& 0.5 & $\co(10^{-7})$ \\
$8\times 10^8$ & $2^{-6}$ & $512 \times 256$ 	& 0.5 & $\co(10^{-7})$ \\
$8\times 10^8$ & $2^{-5}$ & $512 \times 256$ 	& 0.5 & $\co(10^{-7})$ \\
$8\times 10^8$ & $2^{-4}$ & $512 \times 256$ 	& 0.5 & $\co(10^{-7})$ \\
$8\times 10^8$ & $2^{-3}$ & $512 \times 256$ 	& 0.5 & $\co(10^{-7})$ \\
$8\times 10^8$ & $2^{-2}$ & $512 \times 256$ 	& 0.5 & $\co(10^{-7})$ \\
$8\times 10^8$ & $2^{-1}$ & $512 \times 256$ 	& 0.5 & $\co(10^{-7})$ \\
$8\times 10^8$ & $2^{0}$ & $512 \times 256$ 	& 0.5 & $\co(10^{-7})$ \\
$8\times 10^8$ & $2^{1}$ & $512 \times 512$ 	& 0.5 & $\co(4\times 10^{-8})$ \\
$8\times 10^8$ & $2^{2}$ & $512 \times 512$ 	& 0.5 & $\co(4\times 10^{-8})$ \\
$8\times 10^8$ & $2^{3}$ & $512 \times 512$ 	& 0.5 & $\co(4\times 10^{-8})$ \\
$8\times 10^8$ & $2^{4}$ & $512 \times 512$ 	& 0.5 & $\co(4\times 10^{-8})$ \\
$8\times 10^8$ & $2^{5}$ & $512 \times 512$ 	& 0.5 & $\co(4\times 10^{-8})$ \\
$8\times 10^8$ & $2^{6}$ & $1024 \times 512$ & 0.35 & $\co(2\times 10^{-8})$ \\
$8\times 10^8$ & $2^{7}$ & $1024 \times 512$ & 0.35 & $\co(2\times 10^{-8})$ \\
$8\times 10^8$ & $2^{8}$ & $1024 \times 512$ & 0.35 & $\co(2\times 10^{-8})$ \\ \hline
	\end{tabular}
	\caption{(Cont.) List of all 51 simulations discussed in this paper. $\Ra$ is the reference Rayleigh number (see equation \eqref{a44}). Simulations are run with a stiffness parameter $\s$, and with $n_x$ Fourier and $n_z$ Chebyshev basis functions in the horizontal and vertical directions. Timesteps were taken using the Courant-Friedrichs-Lewy (CFL) condition, with a safety factor given in the table. $dt$ is the typical time step, and each simulation was run for at least $\co(1)$ thermal time (about $dt^{-1}$ timesteps). The aspect ratio of all simulations is 2, $\Pr=1$, and $T_t=-20$.}	
	\label{table}
\end{table}

\bibliographystyle{agufull08}
\bibliography{biblio}

\begin{thebibliography}{46}
\providecommand{\natexlab}[1]{#1}
\expandafter\ifx\csname urlstyle\endcsname\relax
  \providecommand{\doi}[1]{doi:\discretionary{}{}{}#1}\else
  \providecommand{\doi}{doi:\discretionary{}{}{}\begingroup
  \urlstyle{rm}\Url}\fi

\bibitem[{\textit{Ahlers et~al.}(2008)\textit{Ahlers, Calzavarini, Araujo,
  Funfschilling, Grossmann, Lohse, and Sugiyama}}]{Ahlers2008}
Ahlers, G., E.~Calzavarini, F.~F. Araujo, D.~Funfschilling, S.~Grossmann,
  D.~Lohse, and K.~Sugiyama (2008), {Non-Oberbeck-Boussinesq effects in
  turbulent thermal convection in ethane close to the critical point},
  \textit{Physical Review E}, \textit{77}(046302), 1--16,
  \doi{10.1103/PhysRevE.77.046302}.

\bibitem[{\textit{Ahlers et~al.}(2009)\textit{Ahlers, Grossmann, and
  Lohse}}]{Ahlers2009}
Ahlers, G., S.~Grossmann, and D.~Lohse (2009), {Heat transfer and large scale
  dynamics in turbulent Rayleigh-B{\'{e}}nard convection}, \textit{Reviews of
  Modern Physics}, \textit{81}(2), 503--537, \doi{10.1103/RevModPhys.81.503}.

\bibitem[{\textit{Alvan et~al.}(2013)\textit{Alvan, Mathis, and
  Decressin}}]{Alvan2013}
Alvan, L., S.~Mathis, and T.~Decressin (2013), {Coupling between internal waves
  and shear-induced turbulence in stellar radiation zones: the critical
  layers}, \textit{Astronomy {\&} Astrophysics}, \textit{553}, A86,
  \doi{10.1051/0004-6361/201321210}.

\bibitem[{\textit{Ansong and Sutherland}(2010)}]{Ansong2010}
Ansong, J.~K., and B.~R. Sutherland (2010), {Internal gravity waves generated
  by convective plumes}, \textit{Journal of Fluid Mechanics}, \textit{648},
  405, \doi{10.1017/S0022112009993193}.

\bibitem[{\textit{Ascher et~al.}(1997)\textit{Ascher, Ruuth, and
  Spiteri}}]{Ascher1997}
Ascher, U.~M., S.~J. Ruuth, and R.~J. Spiteri (1997), {Implicit-explicit
  Runge-Kutta methods for time-dependent partial differential equations},
  \textit{Applied Numerical Mathematics}, \textit{25}(2-3), 151--167,
  \doi{10.1016/S0168-9274(97)00056-1}.

\bibitem[{\textit{Backhaus et~al.}(2011)\textit{Backhaus, Turitsyn, and
  Ecke}}]{Backhaus2011}
Backhaus, S., K.~Turitsyn, and R.~E. Ecke (2011), {Convective instability and
  mass transport of diffusion layers in a Hele-Shaw geometry}, \textit{Physical
  Review Letters}, \textit{106}(10), 104,501,
  \doi{10.1103/PhysRevLett.106.104501}.

\bibitem[{\textit{Baldwin et~al.}(2001)\textit{Baldwin, Gray, Dunkerton,
  Hamilton, Haynes, Randel, Holton, Alexander, Hirota, Horinouchi, Jones,
  Kinnersley, Marquardt, Sato, and Takahashi}}]{Baldwin2001}
Baldwin, M.~P., L.~J. Gray, T.~J. Dunkerton, K.~Hamilton, P.~H. Haynes, W.~J.
  Randel, J.~R. Holton, M.~J. Alexander, I.~Hirota, T.~Horinouchi, D.~B.~A.
  Jones, J.~S. Kinnersley, C.~Marquardt, K.~Sato, and M.~Takahashi (2001), {The
  quasi-biennial oscillation}, \textit{Reviews of Geophysics}, \textit{39}(2),
  179--229.

\bibitem[{\textit{Brummell}(1993)}]{Brummell1993}
Brummell, N.~H. (1993), {The effect of the Prandtl number on penetrative
  convection}, \textit{Geophysical {\&} Astrophysical Fluid Dynamics},
  \textit{68}(1-4), 115--132, \doi{10.1080/03091929308203564}.

\bibitem[{\textit{Brummell et~al.}(2002)\textit{Brummell, Clune, and
  Toomre}}]{Brummell2002}
Brummell, N.~H., T.~L. Clune, and J.~Toomre (2002), {Penetration and
  Overshooting in Turbulent Compressible Convection}, \textit{The Astrophysical
  Journal}, \textit{570}(2), 825--854, \doi{10.1086/339626}.

\bibitem[{\textit{Buffett}(2014)}]{Buffett2014}
Buffett, B. (2014), {Geomagnetic fluctuations reveal stable stratification at
  the top of the Earth's core.}, \textit{Nature}, \textit{507}(7493), 484--487,
  \doi{10.1038/nature13122}.

\bibitem[{\textit{Burns et~al.}((unpublished))\textit{Burns, Vasil, Oishi,
  Lecoanet, Brown, and Quataert}}]{Burns2017}
Burns, K.~J., G.~M. Vasil, J.~S. Oishi, D.~Lecoanet, B.~P. Brown, and
  E.~Quataert ((unpublished)), {Dedalus: A Flexible Pseudo-Spectral Framework
  for Solving Partial Differential Equations}.

\bibitem[{\textit{Dosser and Sutherland}(2011)}]{dosser2011}
Dosser, H.~V., and B.~R. Sutherland (2011), Anelastic internal wave packet
  evolution and stability, \textit{Journal of the Atmospheric Sciences},
  \textit{68}(12), 2844--2859, \doi{10.1175/JAS-D-11-097.1}.

\bibitem[{\textit{Emran and Schumacher}(2008)}]{Emran2008}
Emran, M.~S., and J.~Schumacher (2008), {Fine-scale statistics of temperature
  and its derivatives in convective turbulence}, \textit{Journal of Fluid
  Mechanics}, \textit{611}(2008), 13--34, \doi{10.1017/S0022112008002954}.

\bibitem[{\textit{Garaud et~al.}(2015)\textit{Garaud, Gallet, and
  Bischoff}}]{Garaud2015}
Garaud, P., B.~Gallet, and T.~Bischoff (2015), {The stability of stratified
  spatially periodic shear flows at low P{\'{e}}clet number}, \textit{Physics
  of Fluids}, \textit{27}(8), 084,104, \doi{10.1063/1.4928164}.

\bibitem[{\textit{Goldreich and Kumar}(1990)}]{Goldreich1990}
Goldreich, P., and P.~Kumar (1990), {Wave generation by turbulent convection},
  \textit{The Astrophysical Journal}, \textit{363}, 694, \doi{10.1086/169376}.

\bibitem[{\textit{Grise et~al.}(2010)\textit{Grise, Thompson, and
  Birner}}]{grise2010}
Grise, K.~M., D.~W.~J. Thompson, and T.~Birner (2010), A global survey of
  static stability in the stratosphere and upper troposphere, \textit{Journal
  of Climate}, \textit{23}(9), 2275--2292, \doi{10.1175/2009JCLI3369.1}.

\bibitem[{\textit{Grossmann and Lohse}(2000)}]{Grossmann2000}
Grossmann, S., and D.~Lohse (2000), {Scaling in thermal convection: A unifying
  theory}, \textit{Journal of Fluid Mechanics}, \textit{407}(March), 30,
  \doi{10.1017/S0022112099007545}.

\bibitem[{\textit{Gubbins}(2001)}]{gubbins2001}
Gubbins, D. (2001), The rayleigh number for convection in the earth’s core,
  \textit{Physics of the Earth and Planetary Interiors}, \textit{128}(1), 3 --
  12, \doi{http://dx.doi.org/10.1016/S0031-9201(01)00273-4}, dynamics and
  Magnetic Fields of the Earth's and Planetary Interiors.

\bibitem[{\textit{Hemmati et~al.}(2001)\textit{Hemmati, Moynihan, and {Austen
  Angell}}}]{Hemmati2001}
Hemmati, M., C.~T. Moynihan, and C.~{Austen Angell} (2001), {Interpretation of
  the molten BeF2 viscosity anomaly in terms of a high temperature density
  maximum, and other waterlike features}, \textit{Journal of Chemical Physics},
  \textit{115}(14), 6663--6671, \doi{10.1063/1.1396679}.

\bibitem[{\textit{Hewitt et~al.}(2013)\textit{Hewitt, Neufeld, and
  Lister}}]{Hewitt2013}
Hewitt, D.~R., J.~A. Neufeld, and J.~R. Lister (2013), {Ultimate regime of high
  Rayleigh number convection in a porous medium}, \textit{Physical Review
  Letters}, \textit{108}(224503), 551--586,
  \doi{10.1103/PhysRevLett.108.224503}.

\bibitem[{\textit{Hidalgo et~al.}(2012)\textit{Hidalgo, Fe, Cueto-Felgueroso,
  and Juanes}}]{Hidalgo2012}
Hidalgo, J.~J., J.~Fe, L.~Cueto-Felgueroso, and R.~Juanes (2012), {Scaling of
  convective mixing in porous media}, \textit{Physical Review Letters},
  \textit{109}(26), 1--5, \doi{10.1103/PhysRevLett.109.264503}.

\bibitem[{\textit{Hirose et~al.}(2013)\textit{Hirose, Labrosse, and
  Hernlund}}]{Hirose2013}
Hirose, K., S.~Labrosse, and J.~Hernlund (2013), {Composition and State of the
  Core}, \textit{Annual Review of Earth and Planetary Sciences},
  \textit{41}(1), 657--691, \doi{10.1146/annurev-earth-050212-124007}.

\bibitem[{\textit{Horn et~al.}(2013)\textit{Horn, Shishkina, and
  Wagner}}]{Horn2013}
Horn, S., O.~Shishkina, and C.~Wagner (2013), {On non-Oberbeck – Boussinesq
  effects in three-dimensional Rayleigh – B{\'{e}}nard convection in
  glycerol}, \textit{Journal of Fluid Mechanics}, \textit{724}, 175--202,
  \doi{10.1017/jfm.2013.151}.

\bibitem[{\textit{Hurlburt et~al.}(1994)\textit{Hurlburt, Toomre, Massaguer,
  and Zahn}}]{Hurlburt1994}
Hurlburt, N.~E., J.~Toomre, J.~M. Massaguer, and J.-P. Zahn (1994),
  {Penetration below a convective zone}, \textit{The Astrophysical Journal},
  \textit{421}(January), 245--260.

\bibitem[{\textit{Kerr}(1996)}]{Kerr1996}
Kerr, R.~M. (1996), {Rayleigh number scaling in numerical convection},
  \textit{Journal of Fluid Mechanics}, \textit{310}, 139--179,
  \doi{10.1017/S0022112096001760}.

\bibitem[{\textit{King et~al.}(2012)\textit{King, Stone, Zhang, Gerkema,
  Marder, Scott, and Swinney}}]{king2012}
King, B., M.~Stone, H.~P. Zhang, T.~Gerkema, M.~Marder, R.~B. Scott, and H.~L.
  Swinney (2012), Buoyancy frequency profiles and internal semidiurnal tide
  turning depths in the oceans, \textit{Journal of Geophysical Research:
  Oceans}, \textit{117}(C4), C04,008, \doi{10.1029/2011JC007681}.

\bibitem[{\textit{Large and Andereck}(2014)}]{Large2014}
Large, E., and C.~D. Andereck (2014), {Penetrative Rayleigh-B{\'{e}}nard
  convection in water near its maximum density point}, \textit{Physics of
  Fluids}, \textit{26}(9), 094,101, \doi{10.1063/1.4895063}.

\bibitem[{\textit{{Le Bars} et~al.}(2015)\textit{{Le Bars}, Lecoanet, Perrard,
  Ribeiro, Rodet, Aurnou, and {Le Gal}}}]{Lebars2015}
{Le Bars}, M., D.~Lecoanet, S.~Perrard, A.~Ribeiro, L.~Rodet, J.~M. Aurnou, and
  P.~{Le Gal} (2015), {Experimental study of internal wave generation by
  convection in water}, \textit{Fluid Dynamics Research}, \textit{47}(4),
  045,502, \doi{10.1088/0169-5983/47/4/045502}.

\bibitem[{\textit{Lecoanet and Quataert}(2013)}]{Lecoanet2013}
Lecoanet, D., and E.~Quataert (2013), {Internal gravity wave excitation by
  turbulent convection}, \textit{Monthly Notices of the Royal Astronomical
  Society}, \textit{430}(3), 2363--2376, \doi{10.1093/mnras/stt055}.

\bibitem[{\textit{Lecoanet et~al.}(2015)\textit{Lecoanet, {Le Bars}, Burns,
  Vasil, Brown, Quataert, and Oishi}}]{Lecoanet2015}
Lecoanet, D., M.~{Le Bars}, K.~J. Burns, G.~M. Vasil, B.~P. Brown, E.~Quataert,
  and J.~S. Oishi (2015), {Numerical simulations of internal wave generation by
  convection in water}, \textit{Physical Review E - Statistical, Nonlinear, and
  Soft Matter Physics}, \textit{91}(6), 063,016,
  \doi{10.1103/PhysRevE.91.063016}.

\bibitem[{\textit{Marshall and Schott}(1999)}]{marshall1999}
Marshall, J., and F.~Schott (1999), {Open-ocean convection: Observations,
  theory, and models}, \textit{Reviews of Geophysics}, \textit{37}(1), 1--64,
  \doi{10.1029/98RG02739}.

\bibitem[{\textit{Mellado}(2010)}]{Mellado2010}
Mellado, J.~P. (2010), {The evaporatively driven cloud-top mixing layer},
  \textit{Journal of Fluid Mechanics}, \textit{660}(September), 5--36,
  \doi{10.1017/S0022112010002831}.

\bibitem[{\textit{Moore and Weiss}(1973)}]{Moore1973}
Moore, D.~R., and N.~O. Weiss (1973), {Nonlinear penetrative convection},
  \textit{Journal of Fluid Mechanics}, \textit{61}(03), 553,
  \doi{10.1017/S0022112073000868}.

\bibitem[{\textit{Nycander et~al.}(2015)\textit{Nycander, Hieronymus, and
  Roquet}}]{Nycander2015}
Nycander, J., M.~Hieronymus, and F.~Roquet (2015), {The nonlinear equation of
  state of sea water and the global water mass distribution},
  \textit{Geophysical Research Letters}, \textit{42}, 1--8,
  \doi{10.1002/2015GL065525.1.}

\bibitem[{\textit{Parodi and Emanuel}(2009)}]{parodi2009}
Parodi, A., and K.~Emanuel (2009), A theory for buoyancy and velocity scales in
  deep moist convection, \textit{Journal of the Atmospheric Sciences},
  \textit{66}(11), 3449--3463, \doi{10.1175/2009JAS3103.1}.

\bibitem[{\textit{Pin{\c{c}}on et~al.}(2016)\textit{Pin{\c{c}}on, Belkacem, and
  Goupil}}]{Pincon2016}
Pin{\c{c}}on, C., K.~Belkacem, and M.~J. Goupil (2016), {Astrophysics
  Generation of internal gravity waves by penetrative convection},
  \textit{Astronomy {\&} Astrophysics}, \textit{588}(A122), 1----21.

\bibitem[{\textit{Rogers and Glatzmaier}(2005)}]{Rogers2005}
Rogers, T.~M., and G.~A. Glatzmaier (2005), {Penetrative Convection within the
  Anelastic Approximation}, \textit{The Astrophysical Journal},
  \textit{620}(1), 432--441, \doi{10.1086/423415}.

\bibitem[{\textit{Roquet et~al.}(2015)\textit{Roquet, Madec, Brodeau, and
  Nycander}}]{Roquet2015}
Roquet, F., G.~Madec, L.~Brodeau, and J.~Nycander (2015), {Defining a
  Simplified Yet “Realistic” Equation of State for Seawater},
  \textit{Journal of Physical Oceanography}, \textit{45}(10), 2564--2579,
  \doi{10.1175/JPO-D-15-0080.1}.

\bibitem[{\textit{Shiode et~al.}(2013)\textit{Shiode, Quataert, Cantiello, and
  Bildsten}}]{shiode2013}
Shiode, J.~H., E.~Quataert, M.~Cantiello, and L.~Bildsten (2013), The
  observational signatures of convectively excited gravity modes in
  main-sequence stars, \textit{Monthly Notices of the Royal Astronomical
  Society}, \textit{430}(3), 1736--1745, \doi{10.1093/mnras/sts719}.

\bibitem[{\textit{Siggia}(1994)}]{Siggia1994a}
Siggia, E.~D. (1994), {High Rayleigh Number Convection}, \textit{Annual Review
  of Fluid Mechanics}, \textit{26}(1), 137--168,
  \doi{10.1146/annurev.fluid.26.1.137}.

\bibitem[{\textit{Townsend}(1964)}]{Townsend1964}
Townsend, A.~A. (1964), {Natural convection in water over an ice surface},
  \textit{Q. J. R. Meteorol. Soc.}, \textit{90}(248-59).

\bibitem[{\textit{Urban et~al.}(2012)\textit{Urban, Hanzelka, Kralik, Musilova,
  Srnka, and Skrbek}}]{Urban2012}
Urban, P., P.~Hanzelka, T.~Kralik, V.~Musilova, A.~Srnka, and L.~Skrbek (2012),
  {Effect of boundary layers asymmetry on heat transfer efficiency in turbulent
  Rayleigh-Benard convection at very high Rayleigh numbers}, \textit{Physical
  Review Letters}, \textit{109}(October), 154,301,
  \doi{10.1103/PhysRevLett.109.154301}.

\bibitem[{\textit{Veronis}(1963)}]{Veronis1963}
Veronis, G. (1963), {Penetrative convection}, \textit{The Astrophysical
  Journal}, \textit{137}, 641----663.

\bibitem[{\textit{Vettoretti and Peltier}(2016)}]{Vettoretti2016}
Vettoretti, G., and W.~R. Peltier (2016), {Thermohaline instability and the
  formation of glacial North Atlantic super polynyas at the onset of
  Dansgaard-Oeschger warming events}, \textit{Geophysical Research Letters},
  \textit{43}(10), 5336--5344, \doi{10.1002/2016GL068891}.

\bibitem[{\textit{Vidal and Schaeffer}(2015)}]{Vidal2015}
Vidal, J., and N.~Schaeffer (2015), {Quasi-geostrophic modes in the Earth's
  fluid core with an outer stably stratified layer}, \textit{Geophysical
  Journal International}, \textit{202}(3), 2182--2193,
  \doi{10.1093/gji/ggv282}.

\bibitem[{\textit{Wunsch}(2003)}]{Wunsch2003}
Wunsch, S. (2003), {Stochastic simulations of buoyancy-reversal experiments},
  \textit{Physics of Fluids}, \textit{15}(6), 1442--1456,
  \doi{10.1063/1.1572160}.

\end{thebibliography}
%\bibliography{/home/louis_couston/Dropbox/Bibtex/IRPHE.bib}
%\bibliography{C:/Users/Louis/Dropbox/Bibtex/IRPHE}

\end{document}